\documentclass[conference]{IEEEtran}
\IEEEoverridecommandlockouts
\usepackage{cite}
\usepackage{amsmath,amssymb,amsfonts}
\usepackage{algorithmic}
\usepackage{graphicx}
\usepackage{textcomp}
\usepackage{xcolor}
\usepackage{subcaption}
\usepackage{booktabs}
\usepackage{makecell}
\usepackage{multirow}
\usepackage{url}

\begin{document}

\title{Optimizing Mouse Dynamics for User Authentication by Machine Learning: Addressing Data Sufficiency, Accuracy-Practicality Trade-off, and Model Performance Challenges\\
}

\author{\IEEEauthorblockN{Yi Wang*}
\IEEEauthorblockA{
\textit{The University of Tokyo}\\
Tokyo, Japan \\
yiwangyww@gmail.com}
\and
\IEEEauthorblockN{Chenyv Wu*}
\IEEEauthorblockA{Wuhan, China \\
wcy981021@gmail.com}
\and
\IEEEauthorblockN{Yang Liao}
\IEEEauthorblockA{
\textit{Xi'an Jiaotong Univeristy}\\
Xi'an, China \\
ly905650639@gmail.com}
\and
\IEEEauthorblockN{Maowei You}
\IEEEauthorblockA{
Beijing, China \\
maowei.you@foxmail.com}
}

\maketitle

\begin{abstract}
User authentication is essential to ensure secure access to computer systems, yet traditional methods face limitations in usability, cost, and security. Mouse dynamics authentication, based on the analysis of users' natural interaction behaviors with mouse devices, offers a cost-effective, non-intrusive, and adaptable solution. However, challenges remain in determining the optimal data volume, balancing accuracy and practicality, and effectively capturing temporal behavioral patterns. In this study, we propose a statistical method using Gaussian kernel density estimate (KDE) and Kullback-Leibler (KL) divergence to estimate the sufficient data volume for training authentication models. We introduce the Mouse Authentication Unit (MAU), leveraging Approximate Entropy (ApEn) to optimize segment length for efficient and accurate behavioral representation. Furthermore, we design the Local-Time Mouse Authentication (LT-AMouse) framework, integrating 1D-ResNet for local feature extraction and GRU for modeling long-term temporal dependencies. Taking the Balabit and DFL datasets as examples, we significantly reduced the data scale, particularly by a factor of 10 for the DFL dataset, greatly alleviating the training burden. Additionally, we determined the optimal input recognition unit length for the user authentication system on different datasets based on the slope of Approximate Entropy. Training with imbalanced samples, our model achieved a successful defense AUC $98.52\%$ for blind attack on the DFL dataset and $94.65\%$ on the Balabit dataset, surpassing the current sota performance.
\end{abstract}


\section{Introduction}
User authentication is essential to ensure secure access to computer systems and prevent unauthorized usage\cite{erlich2009authentication}. Traditional authentication methods, such as passwords, auxiliary devices, and biometric recognition, have several limitations. Passwords are susceptible to being guessed, forgotten, or reused across multiple accounts, leading to security vulnerabilities \cite{proctor2002improving}. Auxiliary devices, such as security tokens, can be costly, prone to loss or theft, and add complexity to the user experience. Biometric recognition, such as facial recognition, while more secure, faces challenges such as being tricked by photos or fake videos, high costs, privacy concerns, and varying accuracy due to environmental factors and user appearance changes. In contrast, mouse dynamics, which involves analyzing user behavior through mouse movement patterns, offers a promising alternative. As a means of secondary auxiliary authentication, mouse dynamics authentication is difficult to replicate, protects user privacy, and does not require additional hardware, making it a cost-effective, non-intrusive, and highly adaptable solution. The collection of mouse dynamics data is imperceptible to users and does not disrupt their normal operations or user experience \cite{revett2008survey}. Using the natural interactive behaviors of users, this approach eliminates the need for additional user actions. Consequently, it improves the convenience and fluency of the user experience while simultaneously ensuring system security.

Using computers or other devices equipped with touchpads or mouse input systems, we define authorized users of these devices as legitimate users, while all other individuals are considered unauthorized users. To achieve mouse-dynamics-based authentication, researchers worldwide have invested significant effort and resources into collecting data sets related to mouse dynamics\cite{balabit}, in order to identify user behavior patterns and design user authentication systems. Mouse dynamics datasets typically record behavioral data in time series when users interact with mouse devices. The datasets include cursor positions, kinematic features such as speed and acceleration, and event data such as single-click actions, double-click actions, and scroll wheel events. Given these datasets, existing research has proposed various methods for identifying unauthorized users based on mouse dynamics\cite{fu2020rumba, chong2019user, yao2020identity,antal2019user,antal2020mouse,antal2021sapimouse,yi44trustworthy,quraishi2022secure,siddiqui2021continuous,shen2010user,ahmed2007new,jun2019three,shenchao},. These methods generally involve two steps: first, extracting hand-crafted features from mouse dynamic sequences; second, applying machine learning or deep learning techniques to classify these features for user authentication.

However, existing mouse dynamics-based behavioral authentication systems face the following key challenges:

(\romannumeral1) Determining the appropriate amount of data for effective user authentication remains unresolved. Similar questions have been explored in other fields \cite{Heyman2001dataenough, wortley2005dataenough, kristen2013dataenough}, but not in mouse dynamics authentication. To address this, we propose a method for estimating the required dataset size, avoiding issues of insufficient or excessive data, and providing guidance for experiment design.   

(\romannumeral2)  The length of data segments significantly affects recognition accuracy and real-time performance. Short segments (1–2 seconds) improve responsiveness but lack sufficient behavioral information, reducing accuracy. Longer segments (30 seconds or more) capture richer features but are impractical in scenarios requiring real-time performance.

(\romannumeral3) Mouse dynamics data includes dimensions like time, speed, acceleration, and direction, often noisy and redundant. Traditional models, such as SVMs and Decision Trees, rely on manually extracted features, which are limited to basic statistics and fail to capture complex behavioral patterns.

\begin{figure*}[htbp]
    \centering 
    \includegraphics[width=0.8\linewidth]{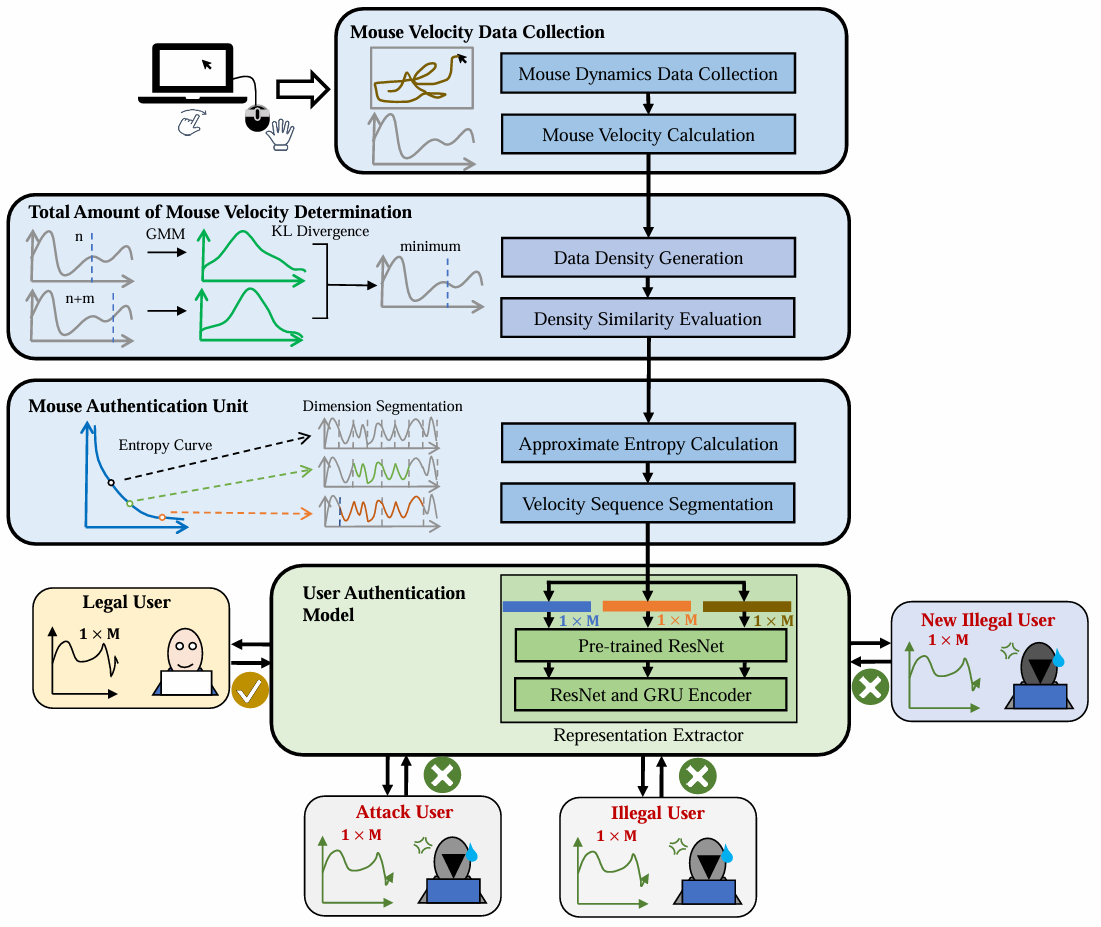} 
    \caption{Structure of the Paper} 
    \label{fig:strcture} 
\end{figure*}

As shown in Figure \ref{fig:strcture}, in this study, we employ statistical methods to address the data volume required for mouse user behavior authentication, aiming to achieve a balance between accuracy and practicality. We propose Local-Time Mouse Authentication (LT-MAuthen), a mouse user verification framework that integrates both local and long-term temporal information.

To construct a user authentication model based on mouse dynamics, we discuss and analyze the raw data content, collection environments, and dataset sizes of different types of mouse dynamics datasets. To ensure adaptability across various devices, reduce model parameter complexity, and enhance inference speed, we compute mouse movement velocity as an input variable for the user authentication system. Furthermore, we propose a general statistical method to determine the appropriate total data volume for modeling user mouse behavior. This method utilizes Gaussian Kernel Density Estimation (KDE) and evaluates the similarity between two density functions with different data volumes using the Kullback–Leibler (KL) divergence. If adding more data results in only minimal changes to the density function—indicated by a KL divergence below a predefined threshold, it suggests that the additional data contributes no new information, and the current data volume is deemed sufficient. 

In practical model training and system deployment, the collected mouse dynamics data must be segmented into multiple short sequences, which serve as inputs for user authentication. We define a Mouse Authentication Unit (MAU) as a smaller, independent temporal sequence segment extracted from continuous mouse trajectory data. Each MAU represents an analyzable behavioral fragment containing sufficient dynamic information to support feature extraction and pattern recognition for user identity verification. To ensure that each MAU encapsulates adequate information without being excessively long—thus compromising system efficiency—we introduce the concept of Approximate Entropy (ApEn) to measure the information content of authentication units. As the length of the MAU increases, approximate entropy decreases while the information content rises, enhancing the discriminability of user behavior. This approach enables a flexible trade-off between authentication speed and recognition accuracy, dynamically determining the optimal MAU length to meet system performance requirements across various application scenarios.

Additionally, to effectively integrate both local and global features of mouse movement sequences, we design a two-step scheme called Local-Time Mouse Authentication. In the first step, a 1D-ResNet is employed to analyze the local features of mouse velocity sequences. In the second step, the trained blocks of the 1D-ResNet are transferred and combined with a GRU (Gated Recurrent Unit) to capture the temporal characteristics within mouse velocity sequences.

Our main contributions are shown in below: 

\begin{enumerate}
    \item We propose a statistical method using KDE and KL divergence to determine the optimal data volume for mouse dynamics authentication model training.
    \item We introduce the Mouse Authentication Unit (MAU) with Approximate Entropy (ApEn) to balance authentication accuracy and efficiency.
    \item We design the LTMouseAuthen framework, combining 1D-ResNet for local feature extraction and GRU for temporal pattern modeling.
\end{enumerate}

The remainder of this paper is organized as follows: Section 2 reviews related work in biometric authentication and mouse dynamics. Section 3 presents our analysis of mouse dynamic data. Sections 4 and 5 detail our methods for determining appropriate data volume and MAU length, respectively. Section 6 describes the proposed LTMouseAuthen framework. Section 7 presents comprehensive experimental results and comparisons. Section 8 discusses limitations and future directions, followed by conclusions in Section 9.

\begin{table*}[ht]
    \centering
    \caption{MAJOR MOUSE DYNAMICS DATASETS}  
    \label{tab:datasetsana}
    \begin{tabular}{c c c c c}
        \toprule
         \textbf{Dataset Name} & \textbf{Time} & \textbf{Content} & \textbf{User Amount} & \textbf{Environment} \\ 
        \midrule 
        \makecell[c]{Mouse-Behavior Data for \\Continuous Authentication~\cite{shenchao2012continuous}} & 2012 & Timestamp, Click, State, ID, X and Y & 28 & Experiment Collection Software \\
        Balabit~\cite{balabit} & 2016 & Timestamp, Click, State, X and Y & 10 & Daily Usage\\ 
        DFL~\cite{antal2019user} & 2018 & Timestamp, Click, State, X and Y & 21 & Experiment Collection Software \\ 
        \makecell[c]{Minecraft-Mouse-Dynamic-\\ Dataset~\cite{siddiqui2021continuous}} & 2022 & Timestamp, Click, Scroll, State, ID, X and Y & 10 & Daily Usage \\
        \bottomrule 
    \end{tabular} 
    \vspace{0.5em}
    \begin{minipage}{0.95\linewidth}
    \end{minipage}
\end{table*}

\section{Related Work}
\textbf{Biometric-based User Authentication} Biometric features like keystroke dynamics, mouse movements, and user-system interactions \cite{Bleha1990ComputerAccess} have been widely used for user authentication. Early work, such as typing patterns \cite{Obaidat1997Verification}, dates back to the 1990s. With the growth of large datasets, machine learning methods have enhanced biometric identification. Bailey \cite{bailey2014user} used a GUI to collect keystroke and mouse dynamics data, applying deep learning for identity verification. Meng \cite{meng2013touch} proposed touch gesture-based authentication on mobile devices, using dynamic training to adapt to data variations. Buriro \cite{buriro2015touchstroke} combined micro-movements, touch strokes, and facial features with random forests and MLPs for classification.

Other biometrics, such as fingerprint recognition \cite{jain2000filterbank} and voice biometrics \cite{kinnunen2010overview}, also offer reliable authentication. Recent studies combine these with behavioral features, like keystroke dynamics and facial recognition, to improve robustness \cite{gupta2015keystroke_face}.

\textbf{User Authentication on Mouse Dynamics} Ahmed et al. \cite{ahmed2007new} first applied machine learning to mouse dynamics, achieving a FAR of 2.46$\%$ and FRR of 2.46$\%$. Shen et al. \cite{shenchao, shenchao2012continuous} used PCA and stream learning to address behavior interference, while Xu et al. \cite{xujian2016mouseauth} applied random forests to reduce overfitting. More recent approaches, such as \cite{jun2019three}, categorize users into groups to reduce authentication time, and \cite{antal2020mouse} used CNNs for deeper feature extraction. Penny et al. \cite{chong2019user} combined CNNs and RNNs for enhanced feature capture. Margit et al. \cite{antal2021sapimouse} released the SapiMouse dataset, and Siddiqui et al. \cite{siddiqui2021continuous} introduced a dataset from Minecraft to avoid dataset homogeneity. However, challenges remain, including the complexity of multidimensional data, non-standardized sequence lengths, and the need for better feature extraction methods.

\section{Analysis of Mouse Dynamic Data}

Mouse dynamics datasets shown in Table \ref{tab:datasetsana} are widely utilized in behavioral biometric authentication as well as in cognitive and psychological research, which are primarily collected from two distinct environments: 

\begin{enumerate}
    \item Daily usage environments
    \item Controlled laboratory environments, with standardized mouse task software
\end{enumerate}

\begin{figure}[h]
  \centering
  \includegraphics[width=\linewidth]{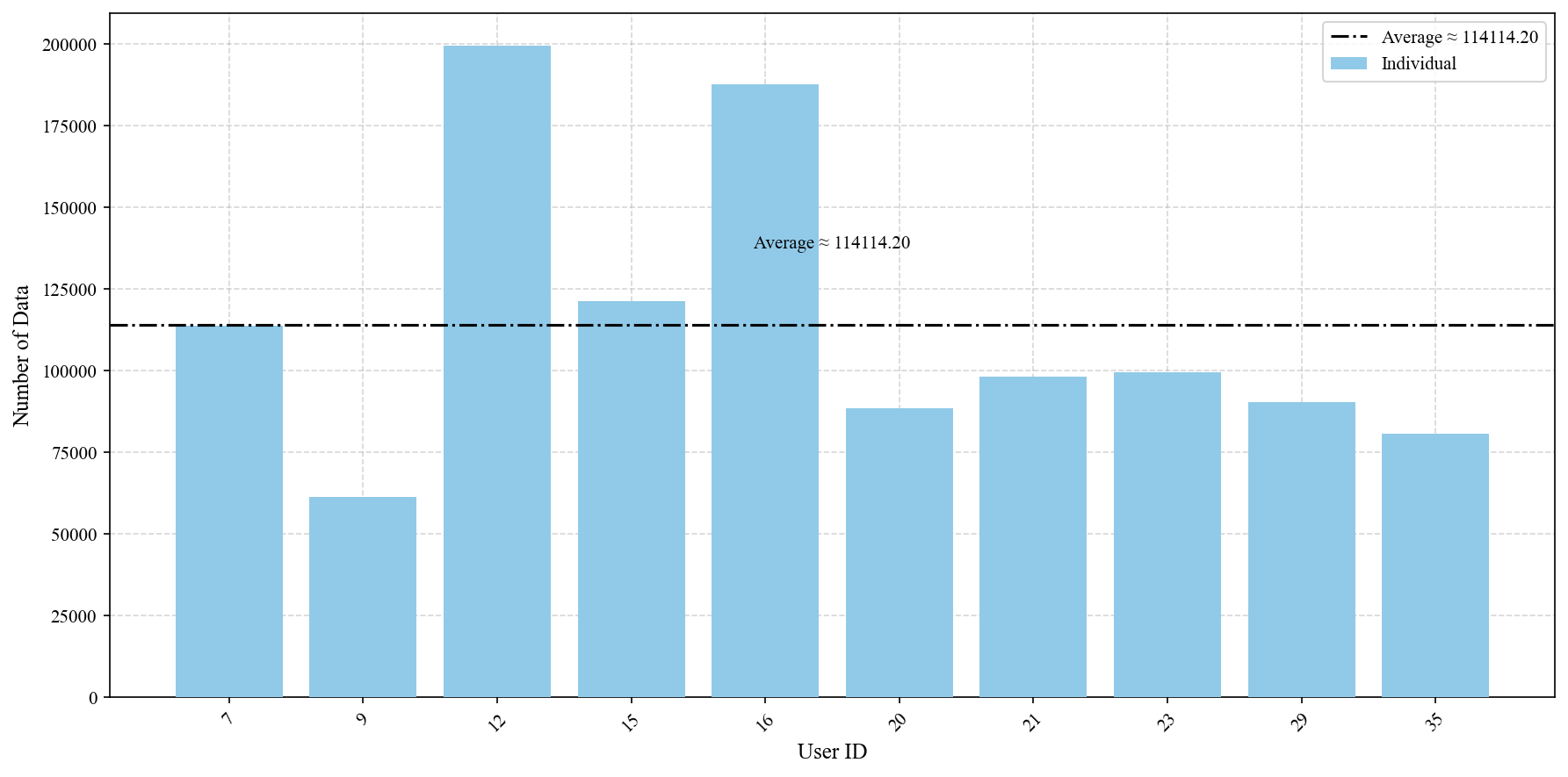}
  \caption{Amount of Individual User Mouse Behavior in the Balabit Dataset}
  \label{fig:balabit_amount}
\end{figure}

\begin{figure}[h]
  \centering
  \includegraphics[width=\linewidth]{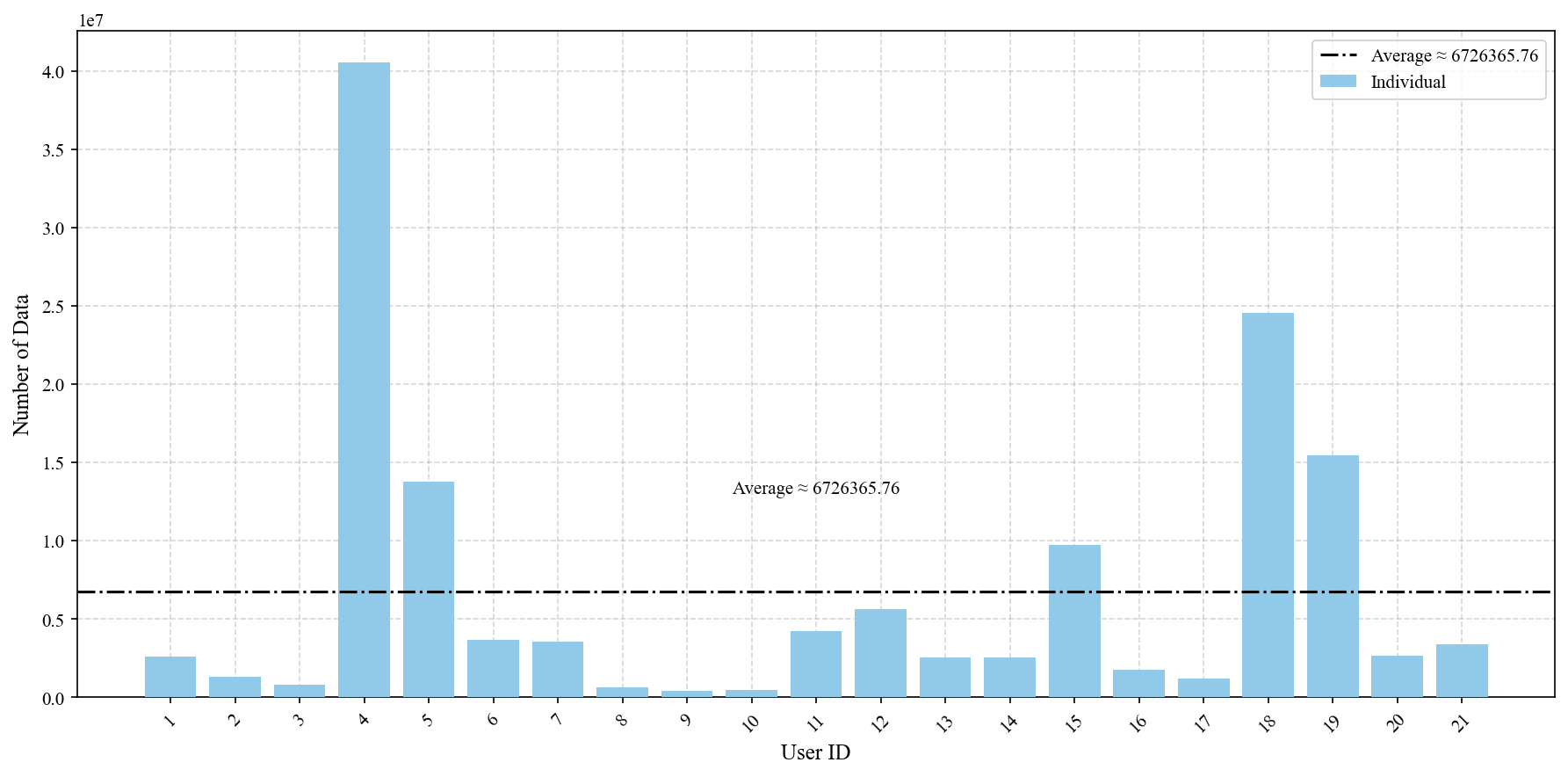} 
  \caption{Amount of Individual User Mouse Behavior in the DFL Dataset}
  \label{fig:dfl_amount}
\end{figure}

Compared to laboratory environments, collecting mouse or touchpad usage data from daily usage environments can more accurately reflect the natural characteristics of user behavior. However, this approach is more costly and time-consuming. As shown in Figure\ref{fig:balabit_amount} and Figure \ref{fig:dfl_amount}, datasets collected from users' daily usage environments, such as the Balabit dataset, are significantly smaller than those obtained from laboratory environments, such as the DFL dataset. To avoid issues of insufficient datasets failing to adequately represent data characteristics, or excessive data leading to wasted time and resources, determining a reasonable data volume is essential. 

Due to differences in application scenarios and data collection methods, the mouse dynamics datasets presented in the Table \ref{tab:datasetsana} have varying feature contents, particularly in the State attribute. For instance, Balabit\cite{balabit} and DFL\cite{antal2019user} datasets' states include dragging, double-click and so on, while Mouse-Behavior Data for Continuous Authentication~\cite{shenchao2012continuous} contains left/right button clicks and movement with left or right button held down, etc. However, all datasets share the basic fundamental movement state. Therefore, before determine a reasonable data volume, we need to choose the variables to describe the mouse dynamic behaviors for our authentication task. To make our method more generalized and reduce the complexity of data processing, we utilize the moving velocity information of user mouse movements as the input data for the authentication system. Furthermore, relying solely on velocity data without directly storing mouse position information enhances user privacy protection and minimizes the risk of sensitive information leakage. 

Given mouse movement trajectory of one specific user as \textbf{$x$} $= \{(x^i,y^i)\}^{N}_{i=1}$, where $x^i$ and $y^i$ represent the mouse coordinates at the $i-$th time point. First, to convert $\{(x^i,y^i)\}^{N}_{i=1}$ to velocity sequence $v = \{v^i\}^{N}_{i=1}$, we calculate the Euclidean distance $d^i$ between adjacent points:

\begin{equation}
    d^i = \sqrt{(x^i - x^{i-1})^2 + (y^i - y^{i-1})^2} 
\end{equation}

Ignoring device latency, the time interval at each time point is fixed and denoted as $\Delta t$. Therefore, the mouse velocity sequence $v^i$ is given by:

\begin{equation}
    v^i = \frac{d^i}{\Delta t}
\end{equation}

\section{Appropriate Volume of Mouse Dynamic Data Determination}
Overly small volume of mouse dynamic data may fail to capture the user's unique operational patterns, while excessively big amount of data may introduce redundant information and high cost. Therefore, determining the optimal volume of mouse dynamics data is crucial for real user authentication task. To establish a unified determination paradigm, we can assume that the user's mouse behavior is largely influenced by the uncertainty caused by the surrounding environment (such as desktop space and current tasks) and the user themselves (such as emotions), but over a long period of use, the mouse dynamics data of individual users will tend to converge, i.e., the user's mouse dynamics data will contain unique and stable identity characteristics.

Based on the this assumption, we estimate the appropriate data quantity by calculating the convergence point of the density function derived from the collected data\cite{7959200}. We use $F(v;n)$ to express the distribution of user$j$ mouse velocity data sequence $v = \{v^i\}^N_{i=1}$, and its density is $f(v;n) = \frac{d}{dx}f(v;n)$ under $n$ time. The density of distribution $f(v;n)$ will vary with different data quantities $n$. If a sufficient amount of data is provided, the observed density will no longer exhibit significant changes with the addition of extra data. For example, if the quantity $\hat{n}$ represents a sufficient amount of mouse dynamics data for user authentication, then the density $p(x,\hat{n}$ should exhibit only minor changes when supplemented with an additional $m$ length of data, i.e.,

\begin{equation}
    f(x;\hat{n}) \approx f(x; \hat{n} + m)
\end{equation}

\begin{figure}[h]
  \centering
  \includegraphics[width=\linewidth]{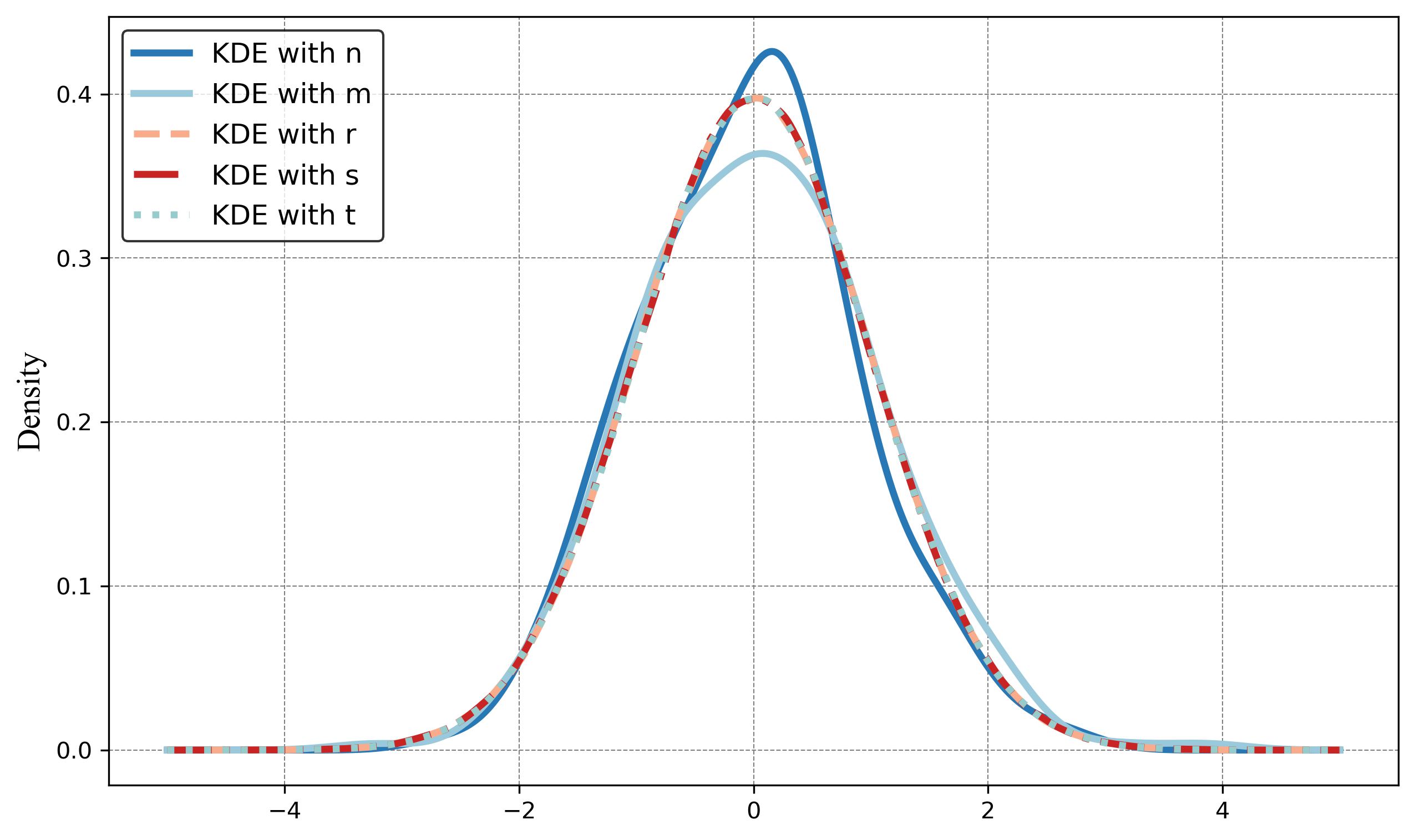} 
  \caption{Illustrationsof the different distribution density with different amount of data for normal distribution $N(0, 1)$ example}
  \label{fig:kdeexam}
\end{figure}

If adding more data does not alter the original data distribution too much, the additional data is considered redundant. Therefore, the observed data quantity may be appropriate because: $\romannumeral1$) This data volume can capture almost all the potential features of user mouse behavior; $\romannumeral2$) Adding more data cannot provide additional useful information but will increase data collection difficulty and computational overhead for the model. As shown in Fig\ref{fig:kdeexam}, We generate random data following a normal distribution $\mathcal{N}(0, 1)$  for specified sample sizes $n, m, r, s$, and $t$, where the intervals are identical, but the sample sizes $r,s$ and $t$ are significantly larger than $n$ and $m$. In the figure\ref{fig:kdeexam}, the distributions for sample sizes $n$ and $m$ are similar, whereas the distributions for sample sizes $r, s$ and $t$ exhibit minimal differences, almost overlapping. It indicates that when the dataset is sufficiently large, adding more data has a negligible impact on altering the overall distribution of the dataset.

We use Gaussian mixture model (GMM)\cite{staffuer1999guassian} to define $f(x;n)$, regarding the complex probability distribution of mouse velocity data as a weighted combination of multiple Gaussian distributions to fit the complex data distribution flexibly. Given a mouse velocity dataset $\{v^i\}^N_{i=1}$ with the density function $f(x;n)$, the density function from data sample $v$ can be estimated by \cite{silverman1986density}

\begin{equation}
    f(x;n) = \frac{1}{n} \sum_{i=1}^{n} \frac{1}{(2\pi)^{\frac{D}{2}} |\Sigma|^{\frac{1}{2}}}
\exp\left(-\frac{1}{2}(v - v^i)^{\top} \Sigma^{-1} (v - v^i)\right)
\label{eq:kde}
\end{equation}

where $D$ is the dimension of the mouse velocity data; $\Sigma$ is the covariance matrix, i.e., bandwidth in 1D data, $|\Sigma|$ is the determinant of bandwidth. In this work, we estimated the bandwidth by\cite{silverman2018density}

\begin{equation}
    h = 1.06 \cdot \hat{\sigma} \cdot n^{-\frac{1}{5}} 
\end{equation}

where $\hat{\sigma}$ is the standard deviation of the dataset $\{v^i\}^N_{i=1}$. 

Thus, we can drive the density function $f(v;n)$ from the mouse velocity data $\{v^i\}^N_{i=1}$. We will assess the similarity between two adjacent kernel functions estimated from $n$ and $n+m$ data observations. Utilizing Kullback-Liebler (KL) divergence \cite{kullback1951KL}, we can evaluate the difference from different density:        

\begin{equation}  
    KL(f(v;n+m)||f(v;n)) = \int \big[f(v;n+m) \times \log \frac{f(v;n+m)}{f(v;n)}\big]
    \label{eq:kldiver}
\end{equation} 

Therefore, the KL can qualify the level of similarity between two density given by mouse velocity data with different length. When $KL(f(v;n+m)||f(v;n))$ approaches 0, it indicates that $f(v;n)$ is extremely close to $f(v;n+m)$, where the additional data would not supply more useful information to density function. Furthermore, it is essential to calculate the KL divergence between $f(v;n+m)$ and $f(v;n+2m)$, as we aim to ensure that the decrease in KL divergence is not abrupt but instead shows a smooth and convergent behavior to ensure our data is enough.

We thus determine the proper amount $n$ of mouse velocity data so that $KL(f(v;n+m)||f(v;n))$ itself is small and change very slightly, even more data added:

\begin{equation}
\begin{cases}
\lvert KL\bigl(f(v;n+m)\|f(v;n)\bigr)\rvert \leq \epsilon_1,\\[6pt]
\lvert KL\bigl(f(v;n+2m)\|f(v;n+m)\bigr)
- KL\bigl(f(v;n+m)\|f(v;n)\bigr)\rvert \leq \epsilon_2
\end{cases}
\label{eq:klconver}
\end{equation}

where $\epsilon$ is a small positive value. It is obvious that a larger value of $\epsilon_1$ and $\epsilon_2$  can lead a small amount of required mouse velocity data.

\section{Mouse Authentication Unit Length Determination}

Having established the amount of data required for the user authentication system to learn mouse movement patterns, we now determine the size of the Mouse Authentication Unit (MAU). Each MAU represents a time-bounded segment of mouse movement data, serving as the foundational element for user verification models. Adjusting the length of the MAU based on each user’s mouse behavior pattern preserves as much valuable information as possible while minimizing the introduction of redundant data, thereby improving the efficiency of the authentication system.

Because information and data predictability are closely linked to data complexity, i.e., entropy, we employ Approximate Entropy (ApEn) \cite{villani2000short} in this study to estimate the information complexity and determine an appropriate MAU length.

Given one individual user's mouse velocity data sequence $\{v^i\}^n_{i=1}$ with certain $n$-dimension, we form a length-$m$ MAU $v(i)$:    

\begin{equation}
    v(i) = (v^i, v^{i+1}, ..., v^{i+m-1}), i = 1,2,...,n-m+1
\end{equation}

Therefore, we create a set $\{v(i)\}_{i=1}^{n-m+1} = \{v(1), v(2), ..., v(n-m+1)\}$ containing all length-$m$ MAU. 



Next, for each pair of length-$m$ MAU $v(p)$ and $v(q)$, we use Chebyshev distance to measure how "close" they are:        

\begin{equation}   
    d[v(p), v(q)] = \max_{s=1,2,\ldots,m} |v^{p+s-1} - v^{q+s-1}|        
\end{equation}   

In ApEn analysis, each length-$m$ MAU $v_j(p)$ is compared against all others $v(q)$ in time series to determine how many of the distances between them lie within a specified tolerance $r$. Formally,     

\begin{equation}  
    C^m_p(r) = \frac{\# \{p \neq q \, | \, d[v(p), v(q)] \leq r\}}{N-m+1}        
\end{equation}  

where $d[·,·]$ denotes the Chebyshev distance metric, and the numerator $\# \{p \neq q \, | \, d[v(p), v(q)] \leq r\}$ counts how many mouse velocity windows remain sufficiently close to $v(p)$ under the threshold $r$. Consequently, $C^m_p(r)$ serves as a measure of the local similarity or “cohesion” for each mouse velocity windows $v(p)$ and forms the basis for evaluating the overall regularity or predictability of the mouse velocity sequence when computing ApEn.    

Then, we repeat the same procedure for another length-$m+1$ MAUs $\{v'(i)\}_{i=1}^{n-m} = \{v(1), v(2), ..., v(n-m)\}$ again. ApEn takes the ratio of these similarity measures at length $m$ and $m+1$ as below:  

\begin{align}
        \text{ApEn}(m) = \frac{1}{n-m+1} \sum_{i=1}^{n-m+1} \log C^m_i(r) \notag \\       
        - \frac{1}{N-m} \sum_{i=1}^{n-m} \log C^{m+1}_i(r)      
\end{align}

where $m$ is length for MAU. The MAU of different lengths possess varying levels of information complexity, which can be quantified by approximate entropy $\text{ApEn}(m)$.  

Approximate entropy decreases as the length of MAU increases, which means it contains more information. But as the length of MAU increase, the race of the increasing of approximate entropy will be slower. It indicates that the data's predictability does not increase much and is insufficient to compensate for the increased collection time required for mouse dynamic sequences. In the subsequent experiments, we aimed to balance the trade-off between the sequence collection time and accuracy. Generally, we selected the length of the sequence with a slow rate of decrease in approximate entropy as the segmentation length for the mouse dynamic sequence.    

\section{Local-Time Mouse Authentication (LTMouseAuthen)}
In order to effectively extract deep features from mouse velocity sequences that can distinguish between different users, we propose a user authentication framework that integrates ResNet residual blocks and a GRU to fully exploit both local and global temporal information. Specifically, the input to the proposed LT-AMouse model is of fixed length. First, a 1D-CNN plus ResNet block module is employed to progressively extract and refine local features. Next, a GRU is utilized to capture contextual correlations among these features. Finally, a fully connected network performs binary classification on the extracted deep features to determine whether the input MAU belongs to the corresponding legitimate user (e.g., user j) or not.

\subsection{ResNet Block for Local Features}
Compared to commonly used multi-modal mouse data (e.g., data with timestamps, (x,y) coordinates, and interaction types), the mouse movement velocity sequence provides only single-channel velocity information. This results in a lower input dimensionality, which accelerates inference and reduces the burden of model deployment. Nevertheless, the reduced dimensionality makes it difficult for conventional manual feature engineering to adequately capture the potential temporal and local detail features. Therefore, we employ a one-dimensional convolutional neural network (1D-CNN) to extract local features from the mouse velocity sequence.      

Specifically, as illustrated in Figure 1, we encode the input N-dimensional mouse velocity sequence using one-dimensional convolution while preserving the sequence length as much as possible to retain its temporal encoding. To further capture deeper local representations of the mouse data, we adopt a ResNet architecture, thereby deepening the convolutional layers and introducing residual connections. This design enables the model to refine key velocity variation patterns while preserving the complete temporal context, laying a solid foundation for subsequent classification or identity verification tasks.      

\subsection{GRU for Time Series Context Information}

\begin{figure*}[htbp]
    \centering 
    \includegraphics[width=0.8\linewidth]{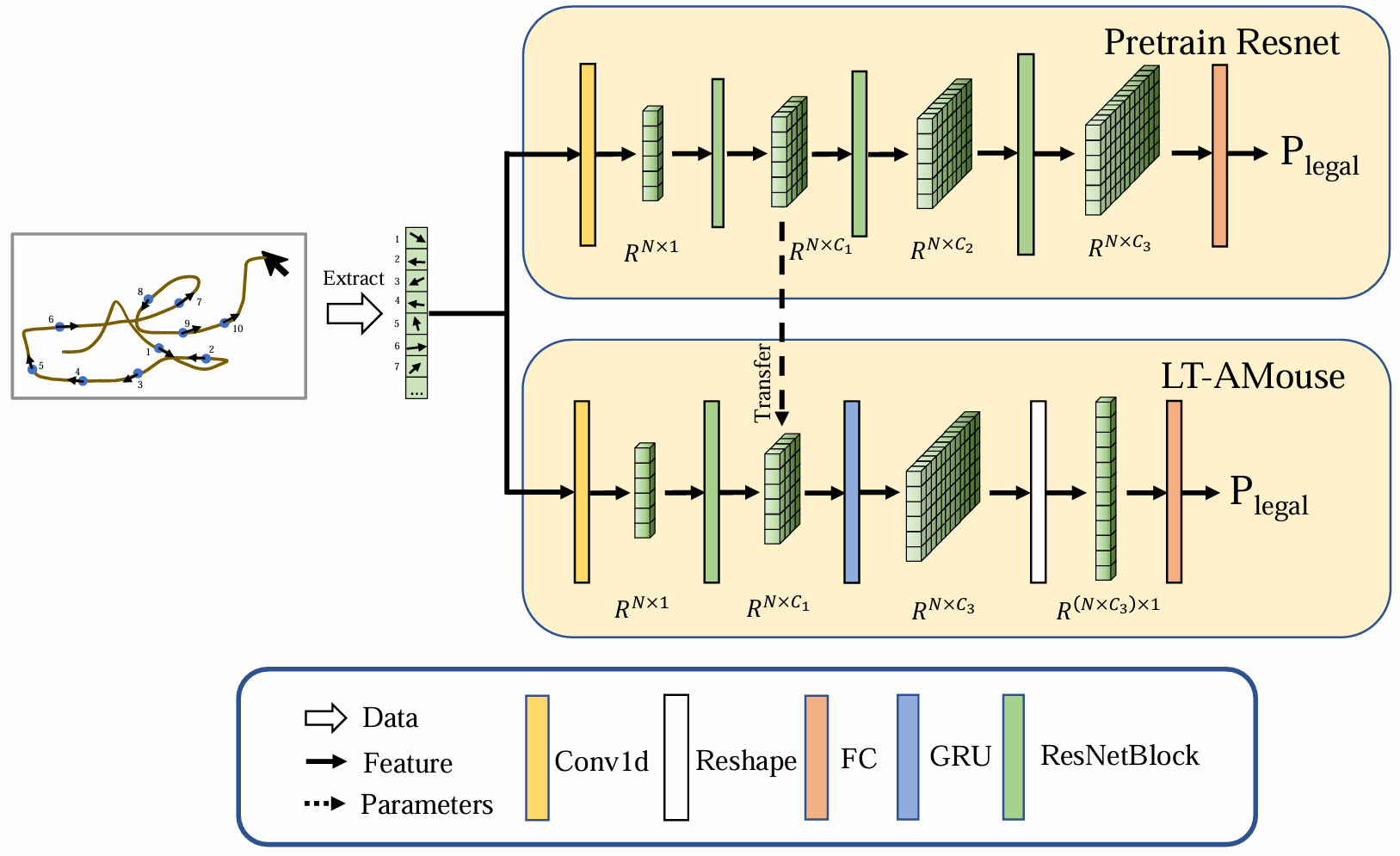} 
    \caption{User Authentication Model} 
    \label{fig:user_authentication_model} 
\end{figure*}

After completing the local convolution and residual encoding, the resulting feature maps primarily emphasize patterns of local velocity changes. To capture global trends and longer-range temporal dependencies, we use a Gated Recurrent Unit (GRU) as the temporal feature extractor. The GRU module can remember past states in the sequence and selectively retain or update crucial information through its gating mechanism, which is particularly important for modeling the temporal evolution of mouse velocity.  

In the overall network architecture, the dimensionality of the features fed into the GRU remains similar with the original sequence length, owing to the ResNet design that preserves sequence length. This allows the GRU to fully utilize the complete temporal information of the original sequence, leading to more effective modeling of velocity patterns.

After extracting both local and global features of the MAU, we feed the resulting hidden states into a fully connected network (FN) for final classification or identity verification. Similar to typical binary classification tasks, this fully connected layer can employ the Softmax function to output a probability distribution, thereby determining whether the input sample belongs to the legitimate user.    

\subsection{Transferring and Training}
During the training process, we formulate user identity verification as a binary classification problem and use cross-entropy loss\cite{Krizhevsky2017ImageNet} as the objective function:    

\begin{equation}
    \mathcal{L}_{C} = - \frac{1}{N}\sum^{C}_{c=1}y_c log \hat{y_{c}}  
\end{equation}
where $N$ is the number of mouse velocity segmentation sequences, $y_c$ is the true value for user authentication, and $\hat{y_{c}}$ is the probability of each velocity segmentation sample after softmax operation.

For the optimization algorithm, we select the Adam optimizer\cite{kingma2014adam} to balance convergence speed and training stability. The parameter $\beta_{1}$ is 0.9 and $\beta_{2}$ is 0.999. 

\section{Experiment}
\subsection{Experimental Setting}
\textbf{Dataset} In this study, the Balabit and DFL datasets were selected as representatives of data collected from daily usage environments and laboratory environments, respectively, to investigate the trade-offs between data sufficiency, efficiency, and accuracy, as well as model performance. We first determine the appropriate amount of data for one individual user to be used as positive samples based on Equation \ref{eq:klconver}, and then randomly select the remaining other users as negative samples to include in the training and testing sets. To verify whether the model is overfitting and to simulate a blind attack scenario, we randomly select unseen users from the remaining samples as unknown samples to be added to the testing set. Considering practical scenarios, user identification should be treated as an imbalanced classification problem, where the amount of positive sample data exceeds that of negative samples. For the DFL dataset, the ratio of positive to negative samples in the training set is 8:1, while for the Balabit dataset, the ratio is 5:1. 

\textbf{Metrics}
We employ the following evaluation metrics to assess the performance of our user authentication system based on mouse velocity sequences and its robustness against attacks:
\begin{enumerate}
    \item \textbf{F1 Score}: The harmonic mean of precision and recall, used to balance the trade-off between false positives and false negatives in imbalanced classification tasks.  
    \item \textbf{Area Under the Curve (AUC)}: The area under the Receiver Operating Characteristic (ROC) curve, which reflects the system's ability to differentiate between legitimate and unauthorized users across varying classification thresholds.
    \item \textbf{Equal Error Rate (ERR)}: The point at which the False Acceptance Rate (FAR) and False Rejection Rate (FRR) are equal, representing a balance between security and usability in the authentication process.
    \item \textbf{Defense Success Rate (DSR)}: A metric used to evaluate the effectiveness of the model in defending against adversarial attacks. It is defined as the percentage of attack attempts that fail to bypass the authentication system.
\end{enumerate}
These metrics collectively ensure a comprehensive evaluation of the system's performance, particularly in the context of imbalanced classification tasks and its robustness under adversarial conditions.


\subsection{Proper Volume of Data}

\begin{figure*}[htbp]
    \centering
    \begin{subfigure}[b]{0.48\textwidth}
        \centering
        \includegraphics[width=\textwidth]{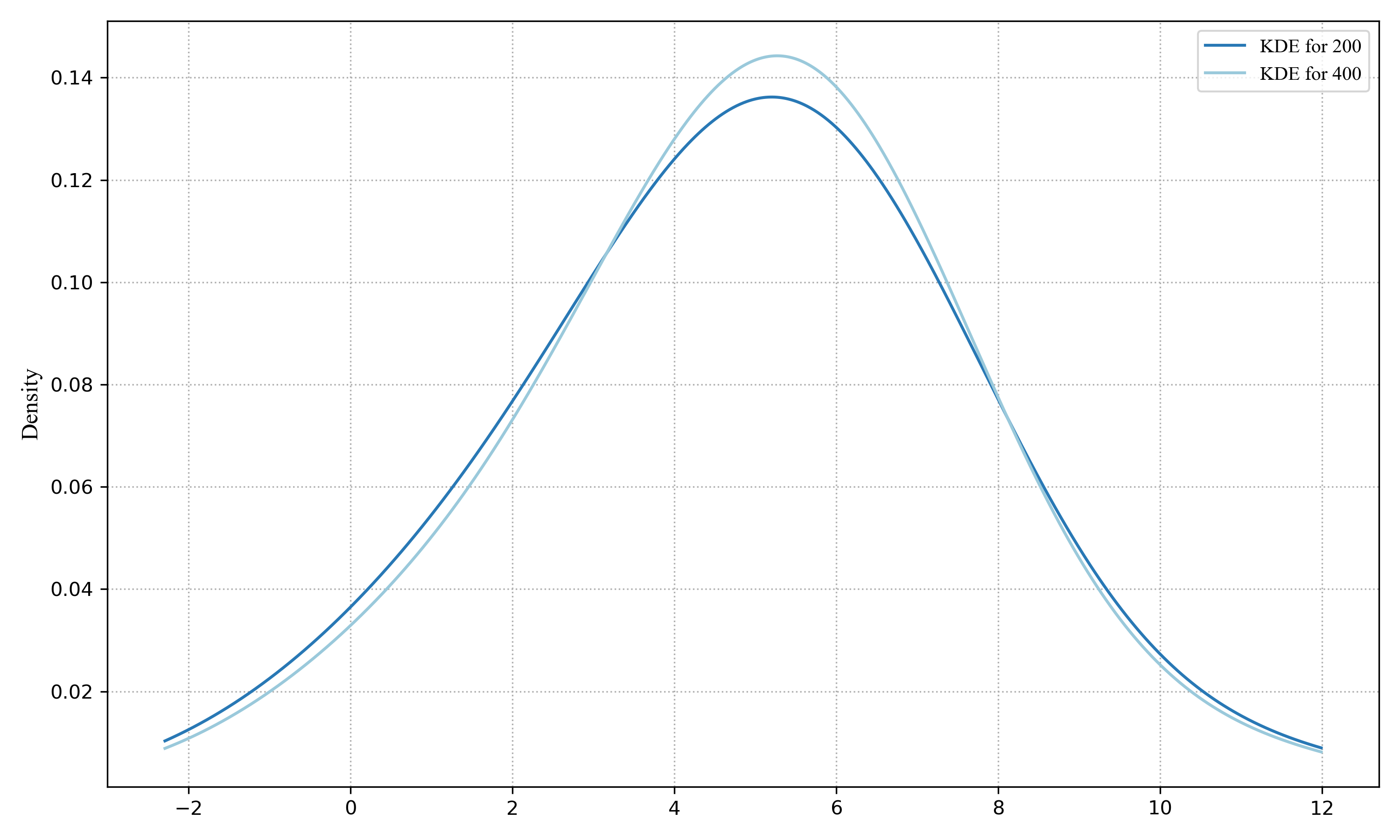}
        \caption{KDE of Small U12}
        \label{fig:kde_200_400}
    \end{subfigure}
    \quad
    \begin{subfigure}[b]{0.48\textwidth}
        \centering
        \includegraphics[width=\textwidth]{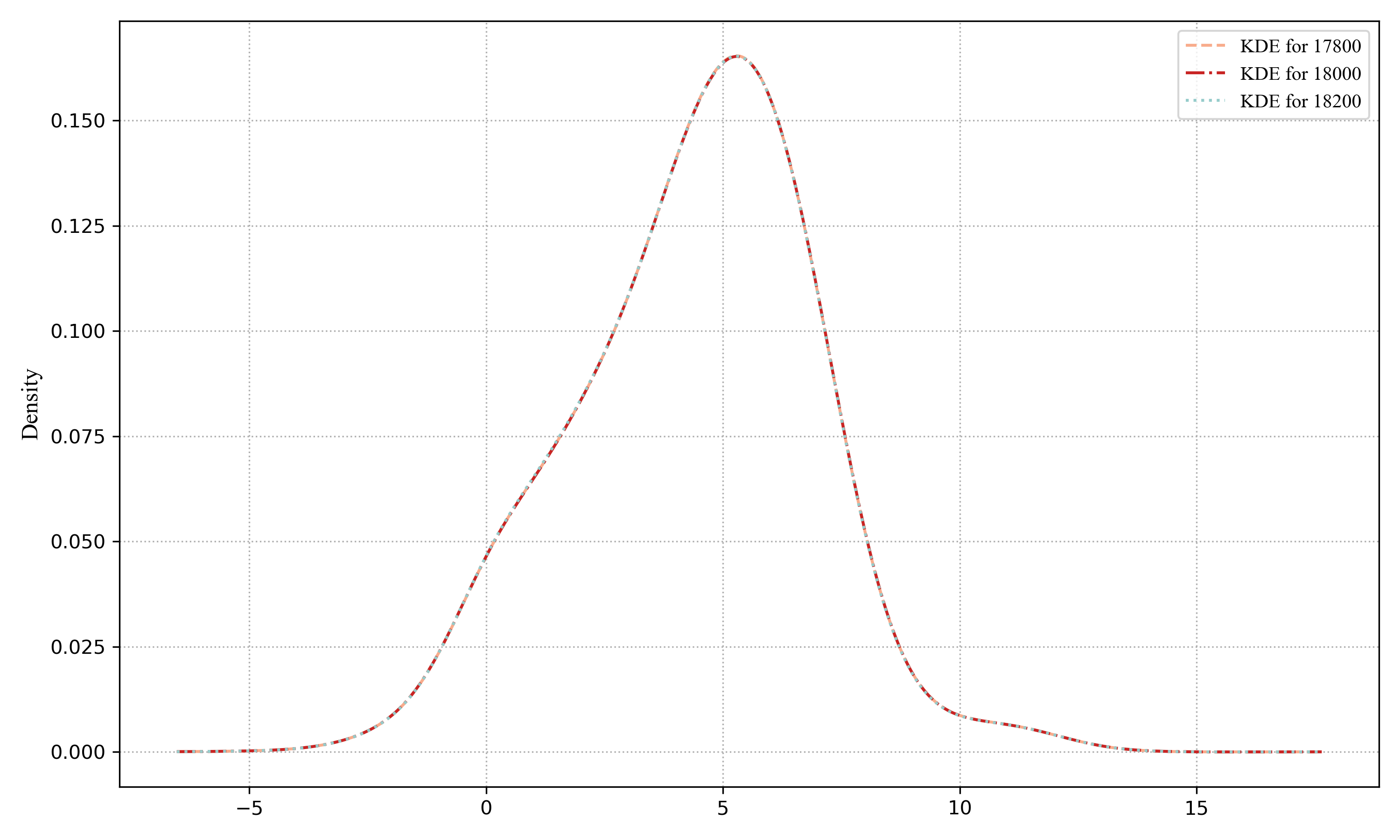}
        \caption{KDE of Enough U12}
        \label{fig:kde_17899_18200}
    \end{subfigure}

    \vspace{1em}

    \begin{subfigure}[b]{0.48\textwidth}
        \centering
        \includegraphics[width=\textwidth]{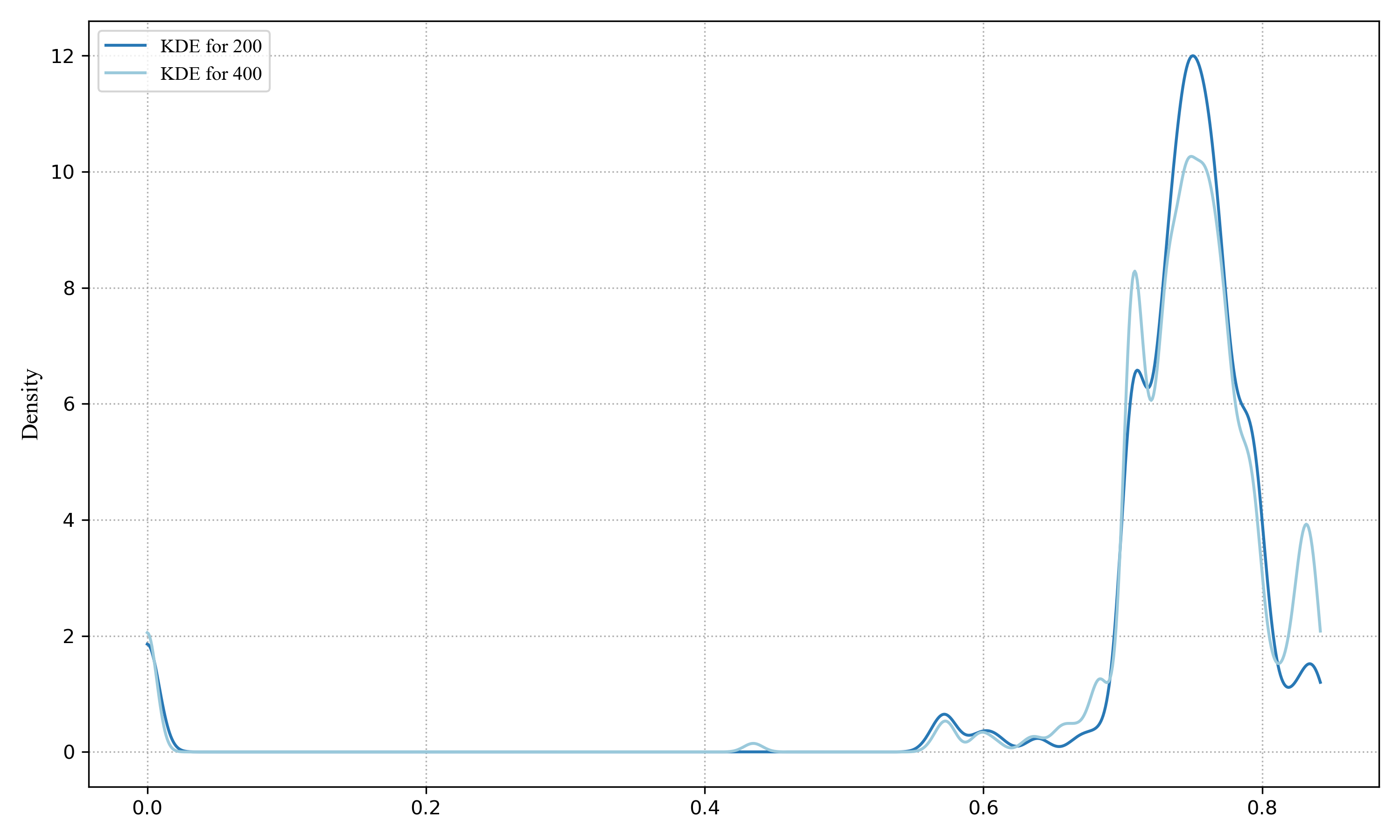}
        \caption{KDE of Small U19}
        \label{fig:dfl_before}
    \end{subfigure}
    \quad
    \begin{subfigure}[b]{0.48\textwidth}
        \centering
        \includegraphics[width=\textwidth]{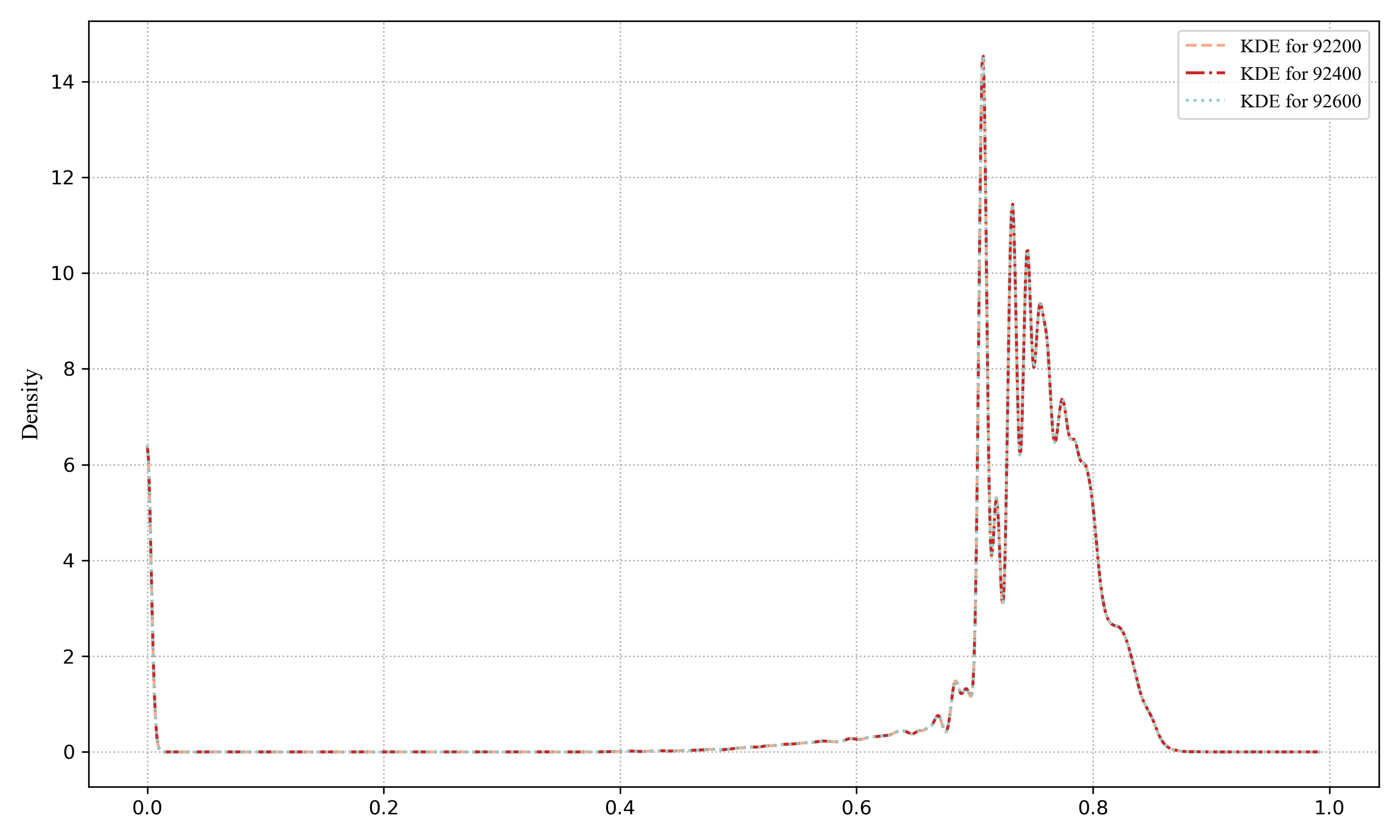}
        \caption{KDE of Enough U19}
        \label{fig:dfl_after}
    \end{subfigure}

    \caption{Comparison of KDE of Different Volume of Mouse Velocity Data for Balabit and DFL Dataset Example}
    \label{fig:kde_change}
\end{figure*}

In this study, mouse velocity sequences were extracted from two mouse dynamics datasets, Balabit and DFL mentioned in Section 3. Each user in the Balabit and DFL datasets is associated with multiple CSV files, each representing a distinct session unit. There is a clear discontinuity between the ending timestamp of one CSV file and the starting and ending timestamps of the subsequent file. Considering the potential variations in mouse operation habits across different time periods, which may result in differing data distributions, we first calculate the optimal data quantity for each individual CSV file and then sum these quantities to obtain the total data volume.

After calculating the mouse velocity sequences, due to the dataset being divided into multiple CSV files with clear temporal discontinuities, we computed the velocity sequences for each CSV file. To calculate the optimal data volume for a single session for each user, based on Equation\ref{eq:kde}, we first generate KDEs for different data volumes within each session, using a step size of 200. As shown in Figure \ref{fig:kde_change}, taking User 12 in Balabit and User 19 in DFL as example, the kernel density changes significantly with increasing data when the mouse velocity data volume is relatively small. At data volumes of $n = 200$ and $n = 400$, the kernel densities for two velocity sequence lengths differ markedly. However, when the data volume is large, even substantial data increases result in minimal kernel density differences. For example, when the data volume reaches sufficient value, mouse velocity data captures all variations, and the kernel density is not significantly different around it.  

To better define the differences between kernel densities, we calculate the KL divergence between kernel densities represented by data of two different sample sizes using Equation\ref{eq:kldiver}. Additionally, we use Equation\ref{eq:klconver} to determine when the KL divergence converges to a sufficiently small value, indicating that the distributions of the two datasets exhibit only minimal changes. If the two thresholds in Equation\ref{eq:klconver} are set to larger values, it results in smaller data volume; conversely, smaller thresholds yield larger data volume. To obtain conservative results, we set $\epsilon_1$ to $1 \times 10^{-4}$ in this study. For the Balabit dataset, considering the complexity of mouse dynamics in real-world scenarios, we adopt a more conservative threshold $\epsilon_2$ as $1 \times 10 ^{-7}$; for the DFL dataset, we adopted a more aggressive strategy, setting $\epsilon_2$ as $1 \times 10 ^{-6}$.          

Taking User 12 from the Balabit dataset and User 9 from the DFL dataset as examples, both of which have relatively large original data volumes, we present the variation trends of the KL divergence values corresponding to Equation\ref{eq:kldiver} across different session data volumes. As illustrated in Figure \ref{fig:user9_kl}, the KL divergence decreases and stabilizes as the data volume grows, suggesting diminishing distributional differences between the mouse dynamic datasets with different volume. Furthermore, adding more mouse velocity data contributes minimal additional information, indicating saturation in the data's informational content.  

\begin{figure}[htbp] 
  \centering
  \includegraphics[width=\linewidth]{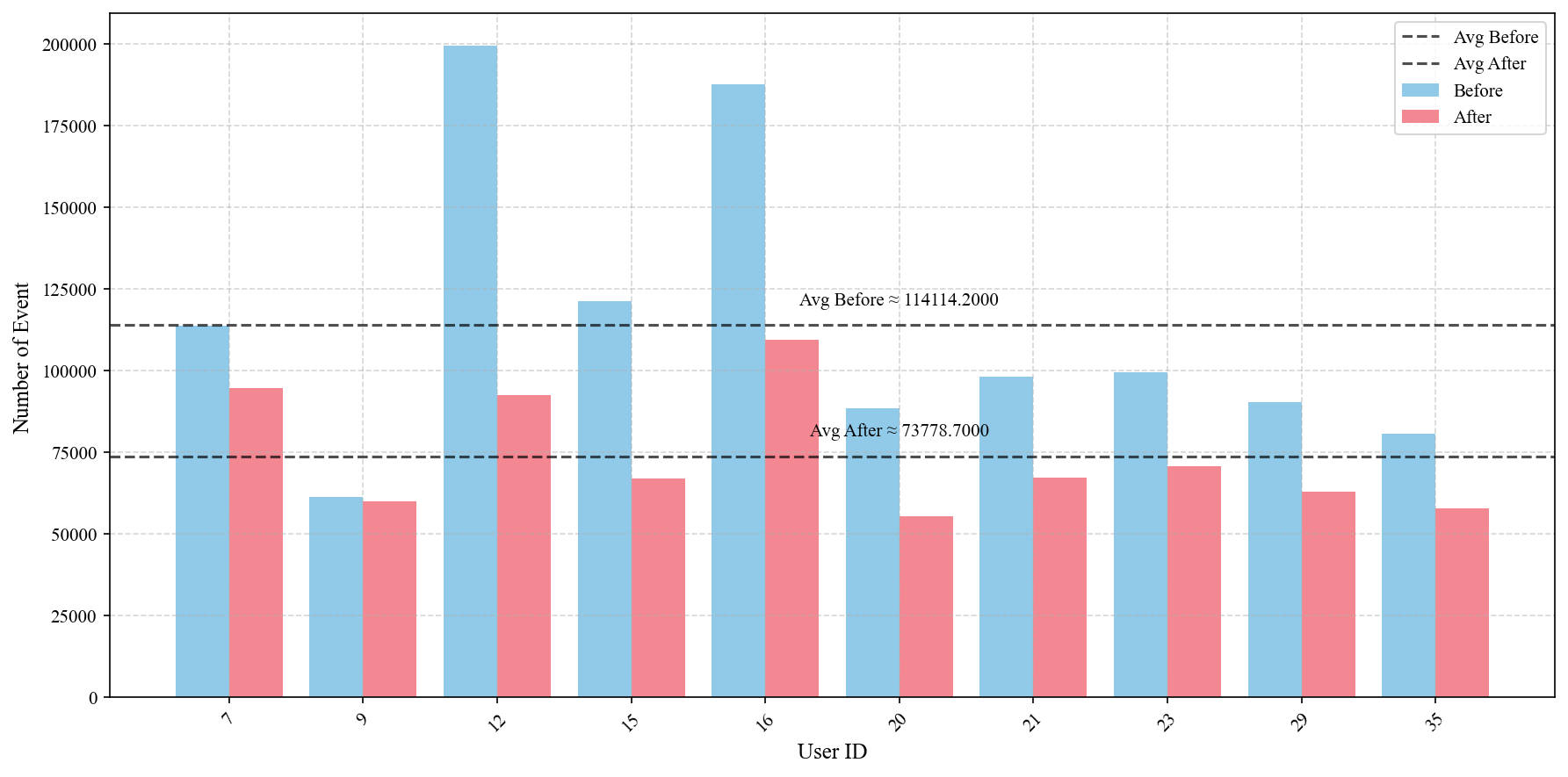}        
  \caption{Proper and Total Volume of Balabit Dataset}  
  \label{fig:balabit_compare}
\end{figure}

\begin{figure}[htbp]
  \centering
  \includegraphics[width=\linewidth]{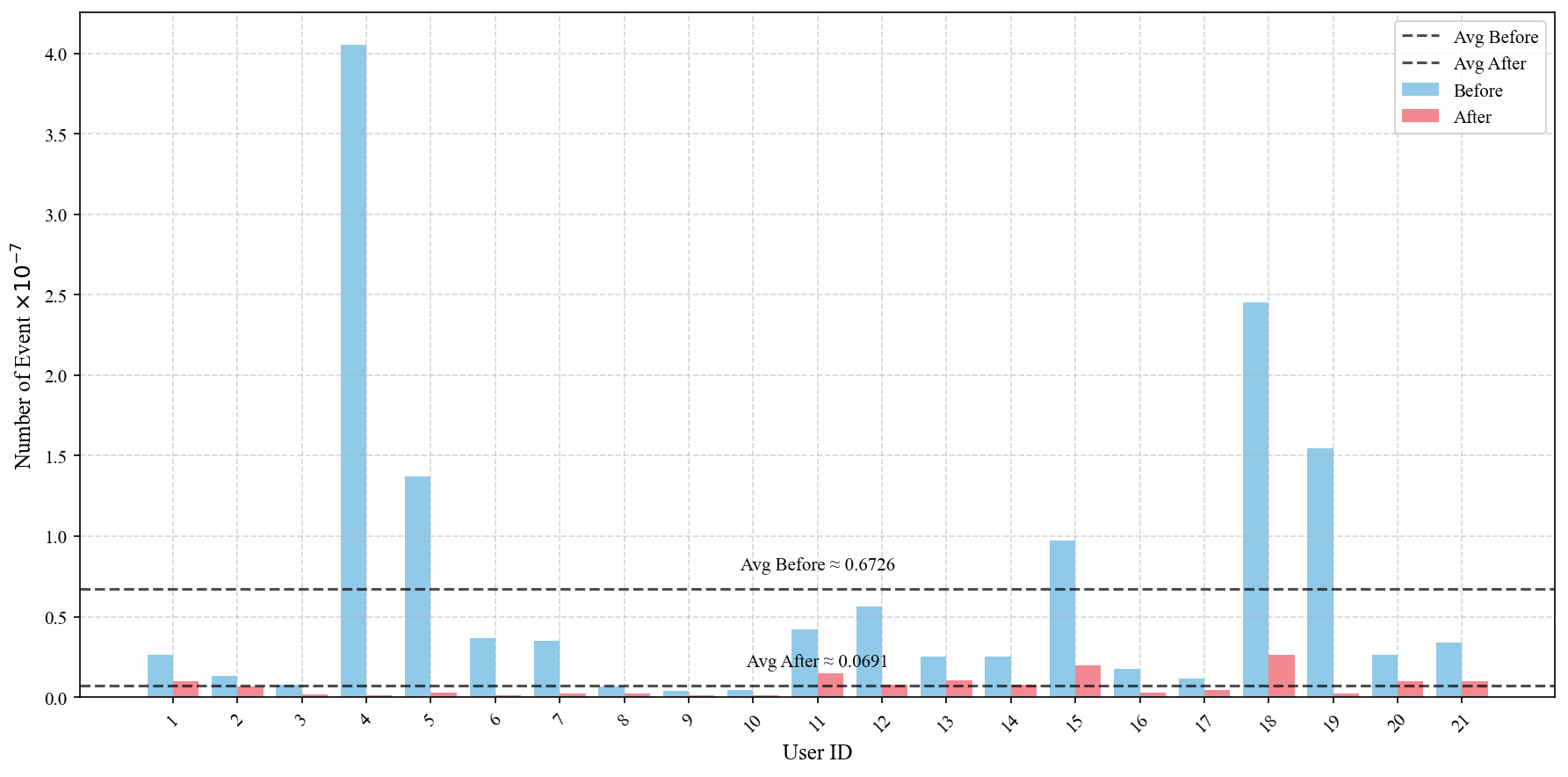}  
  \caption{Proper and Total Volume of DFL Dataset}
  \label{fig:dflcompare}
\end{figure}

\begin{figure*}[htbp]
    \centering
    \begin{subfigure}[b]{0.23\textwidth}  
        \includegraphics[width=\textwidth]{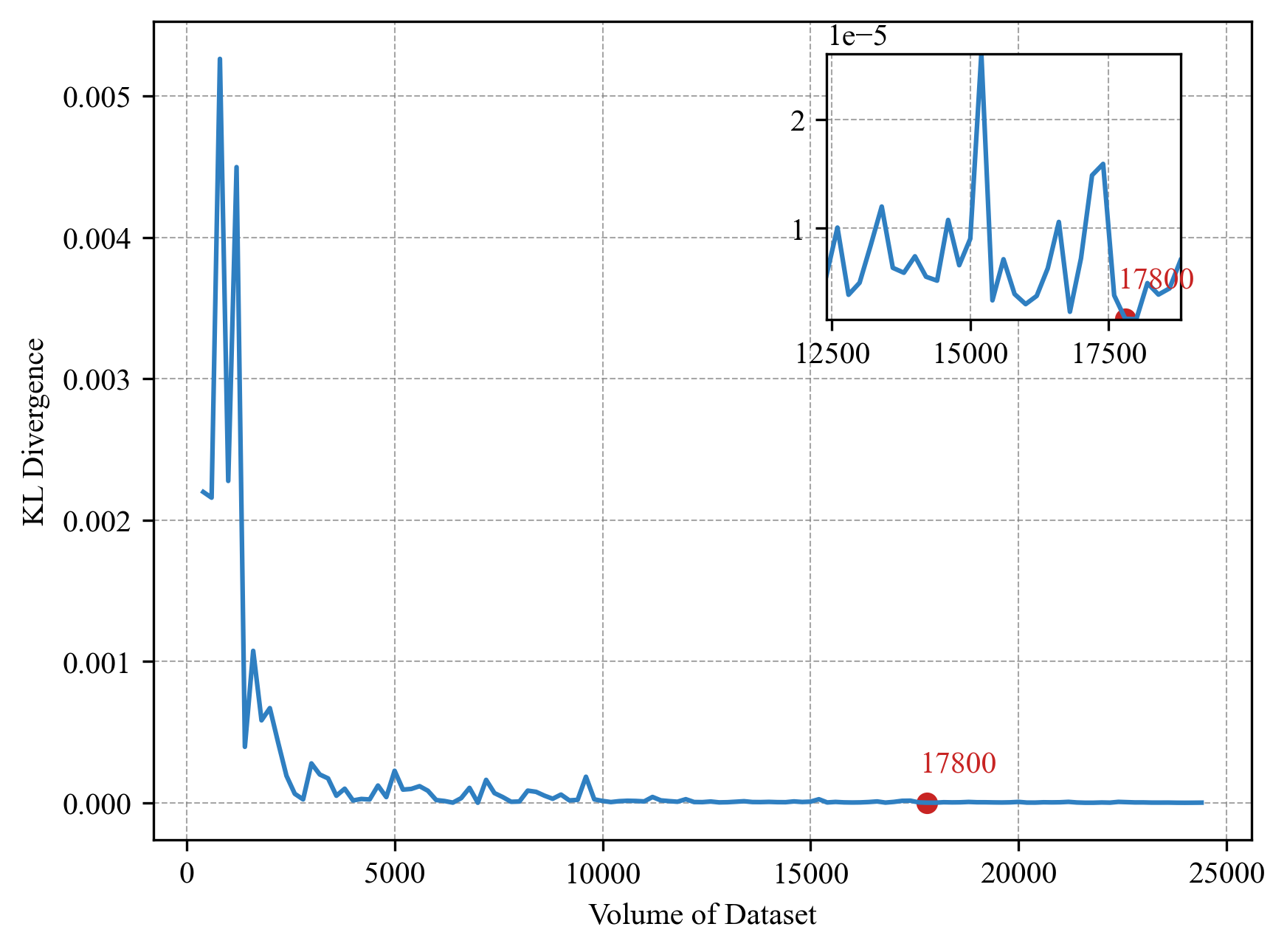}
        \caption{Balabit User12 Session1}
    \end{subfigure}
    \hfill
    \begin{subfigure}[b]{0.23\textwidth}
        \includegraphics[width=\textwidth]{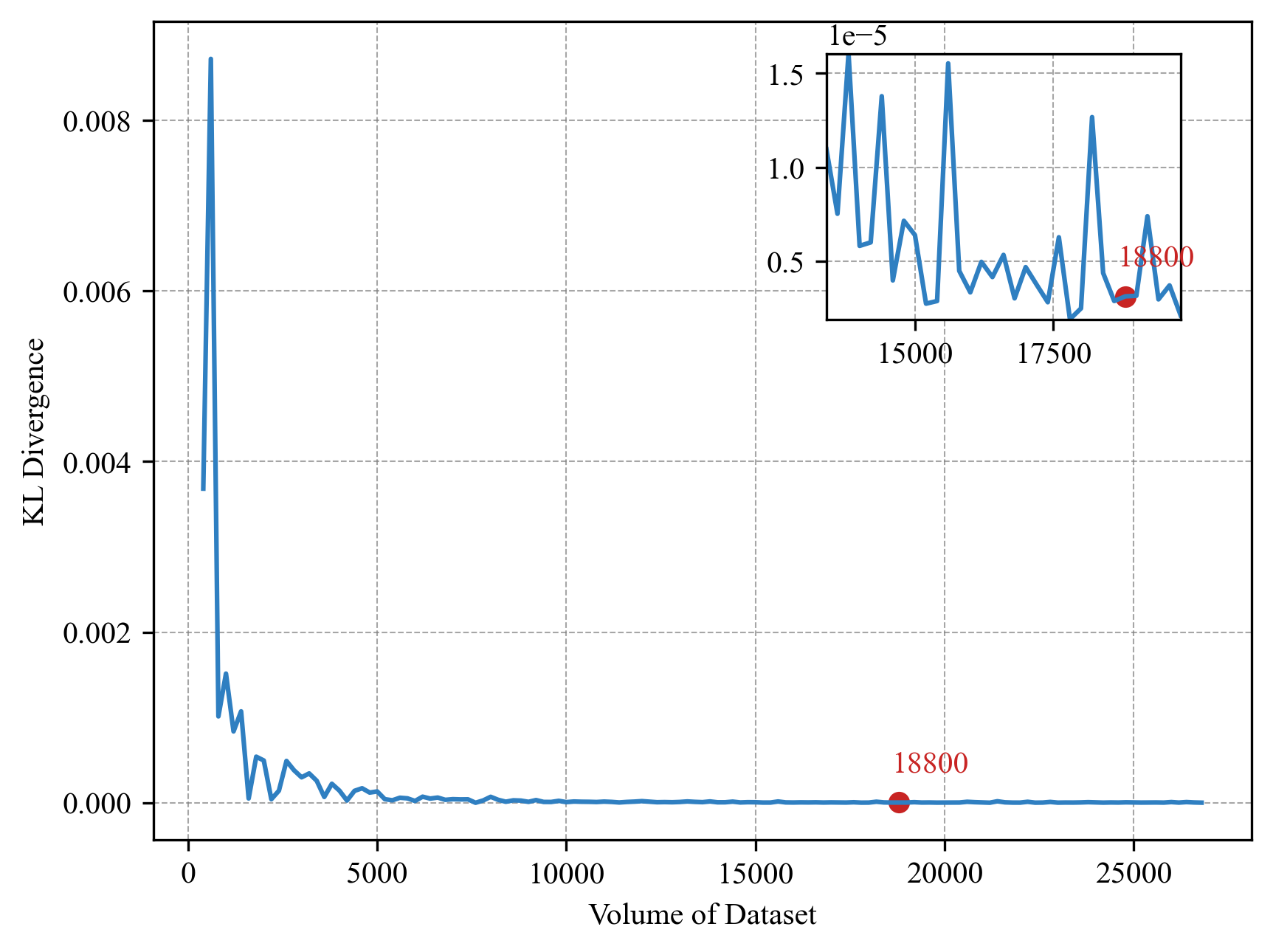}
        \caption{Balabit User12 Session2}
    \end{subfigure}
    \hfill
    \begin{subfigure}[b]{0.23\textwidth}
        \includegraphics[width=\textwidth]{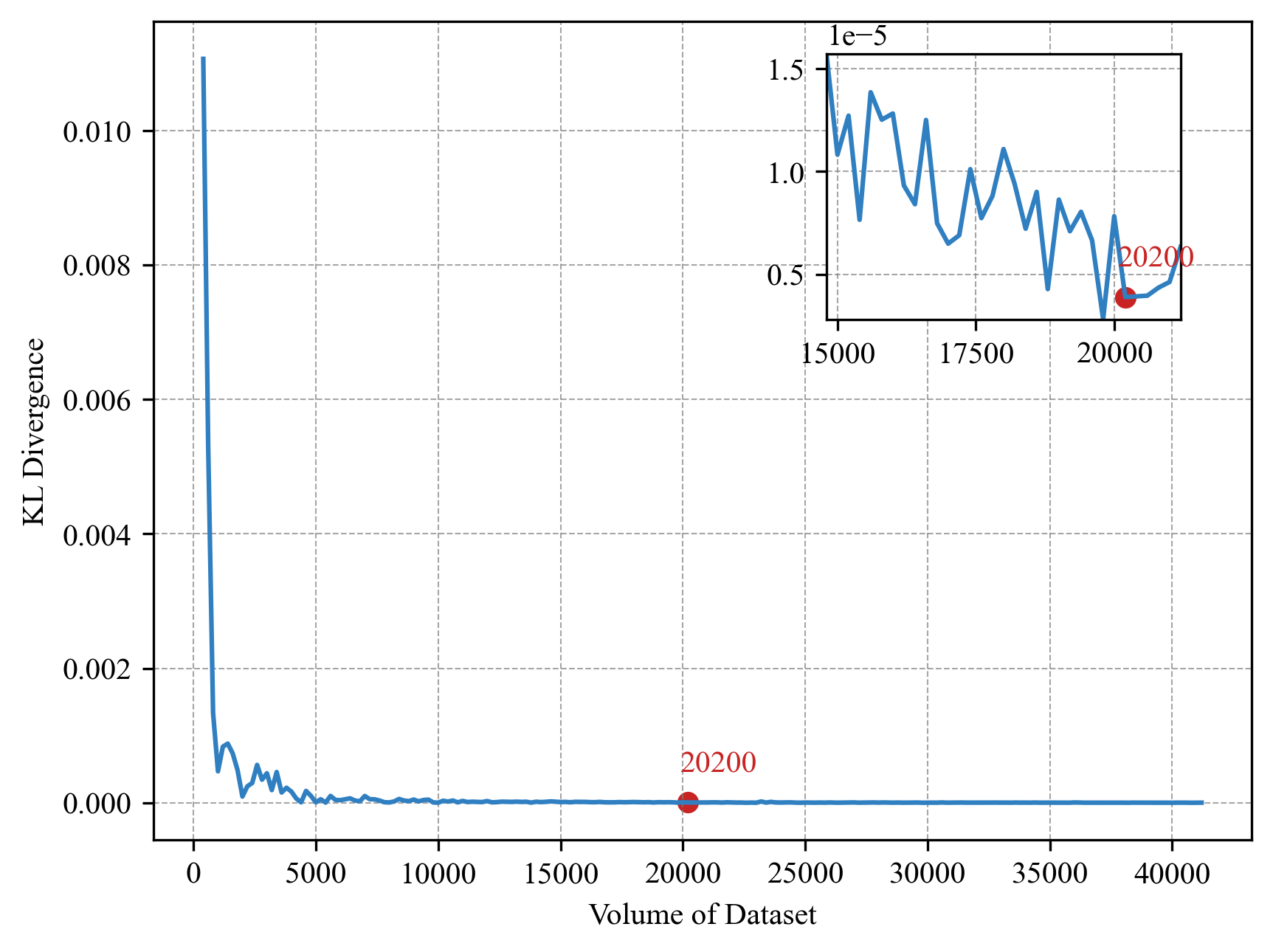}
        \caption{Balabit User12 Session3}
    \end{subfigure}
    \hfill
    \begin{subfigure}[b]{0.23\textwidth}
        \includegraphics[width=\textwidth]{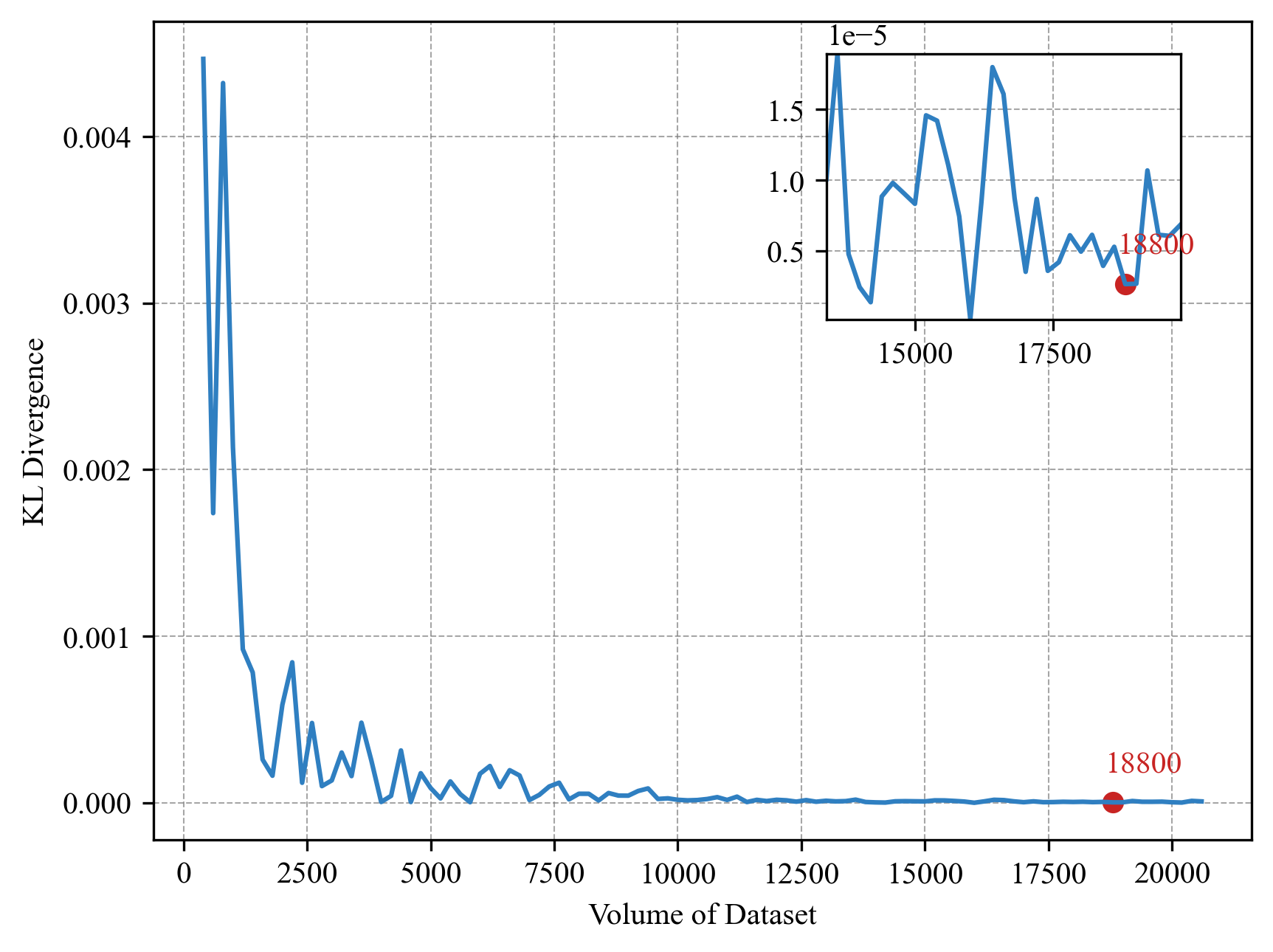}
        \caption{Balabit User12 Session4}
    \end{subfigure}
    
    \vspace{2mm}
    
    \begin{subfigure}[b]{0.23\textwidth}
        \includegraphics[width=\textwidth]{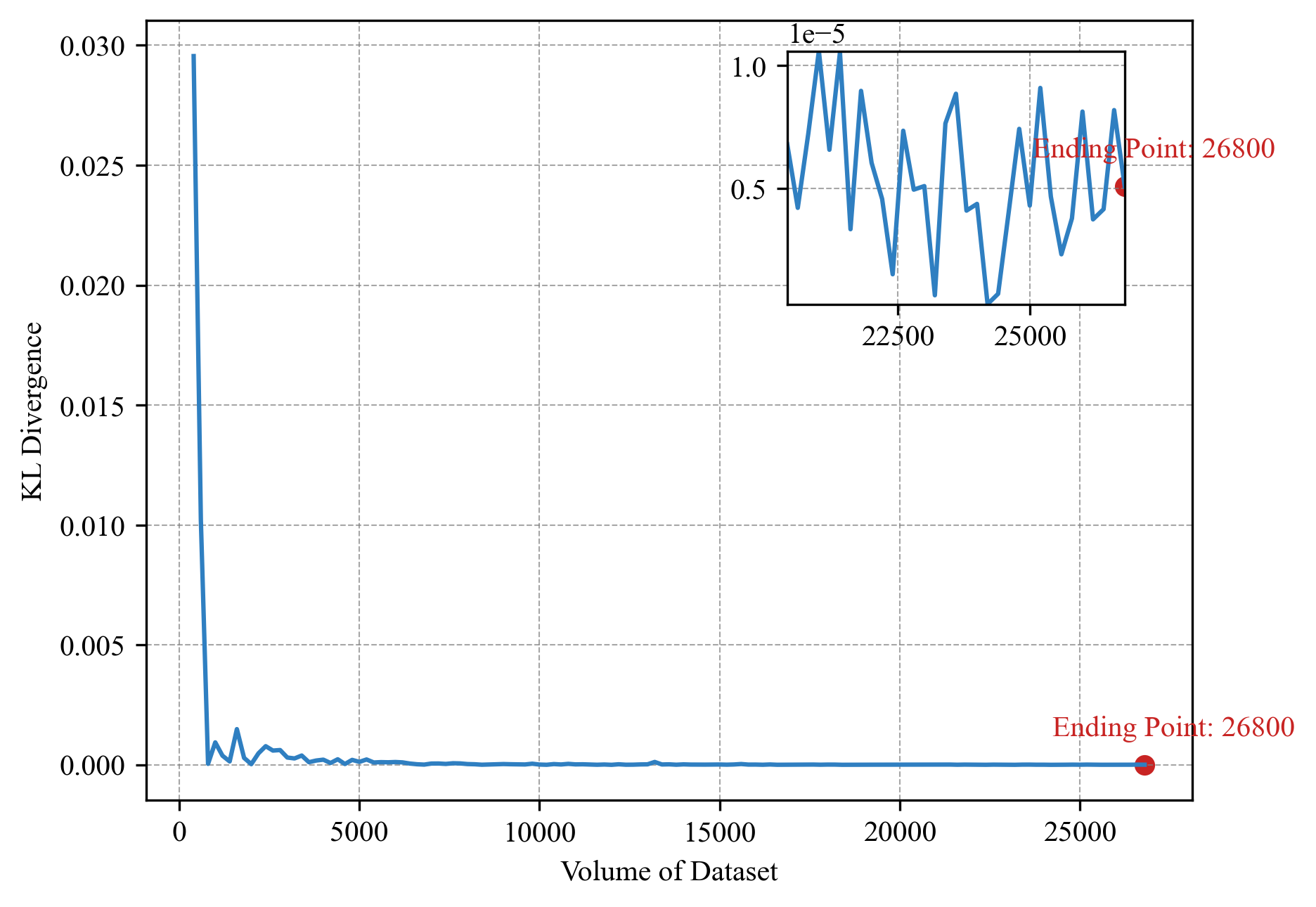}
        \caption{Balabit User12 Session5}
    \end{subfigure}
    \hfill
    \begin{subfigure}[b]{0.23\textwidth}
        \includegraphics[width=\textwidth]{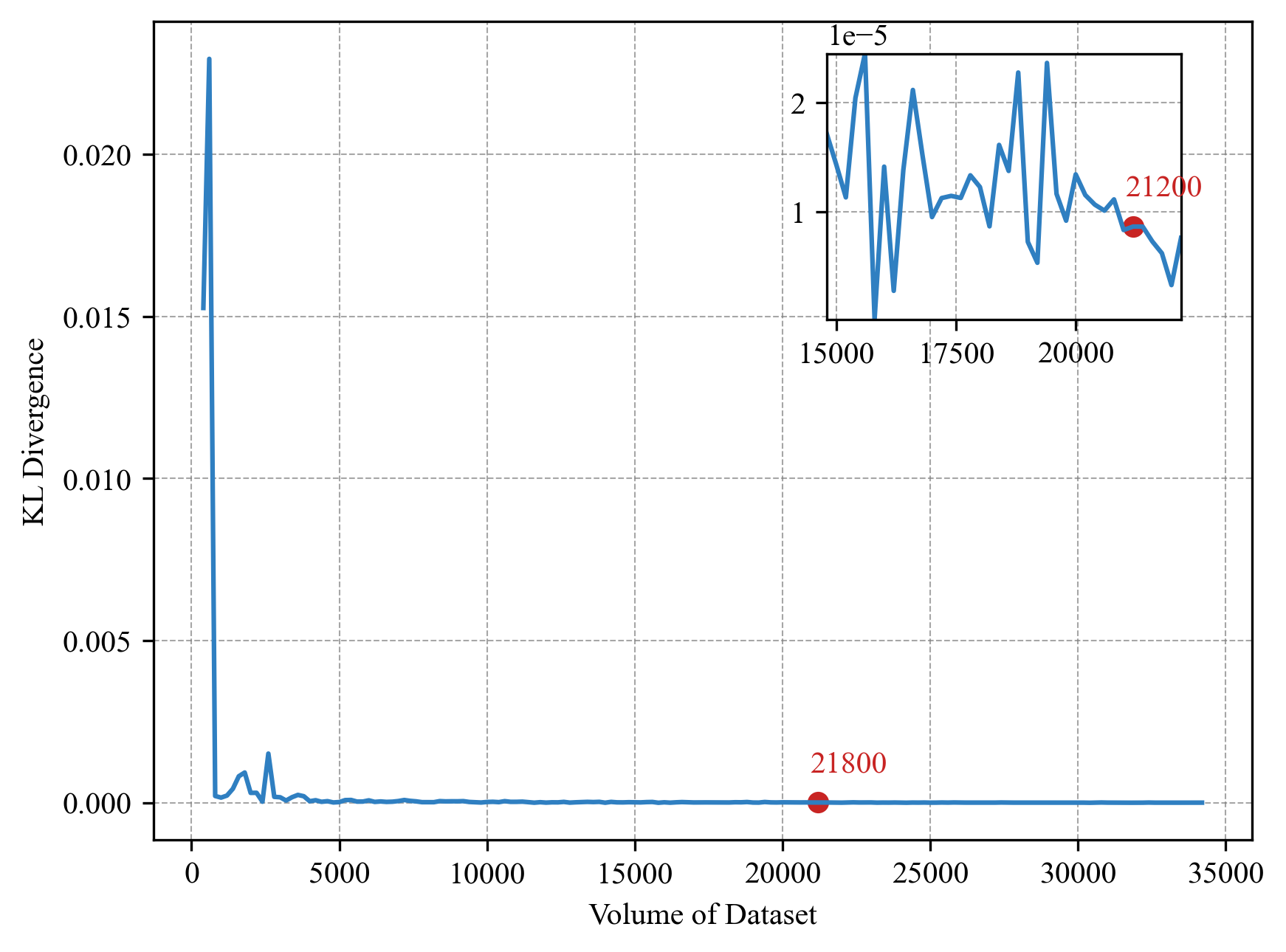}
        \caption{Balabit User12 Session6}
    \end{subfigure}
    \hfill
    \begin{subfigure}[b]{0.23\textwidth}
        \includegraphics[width=\textwidth]{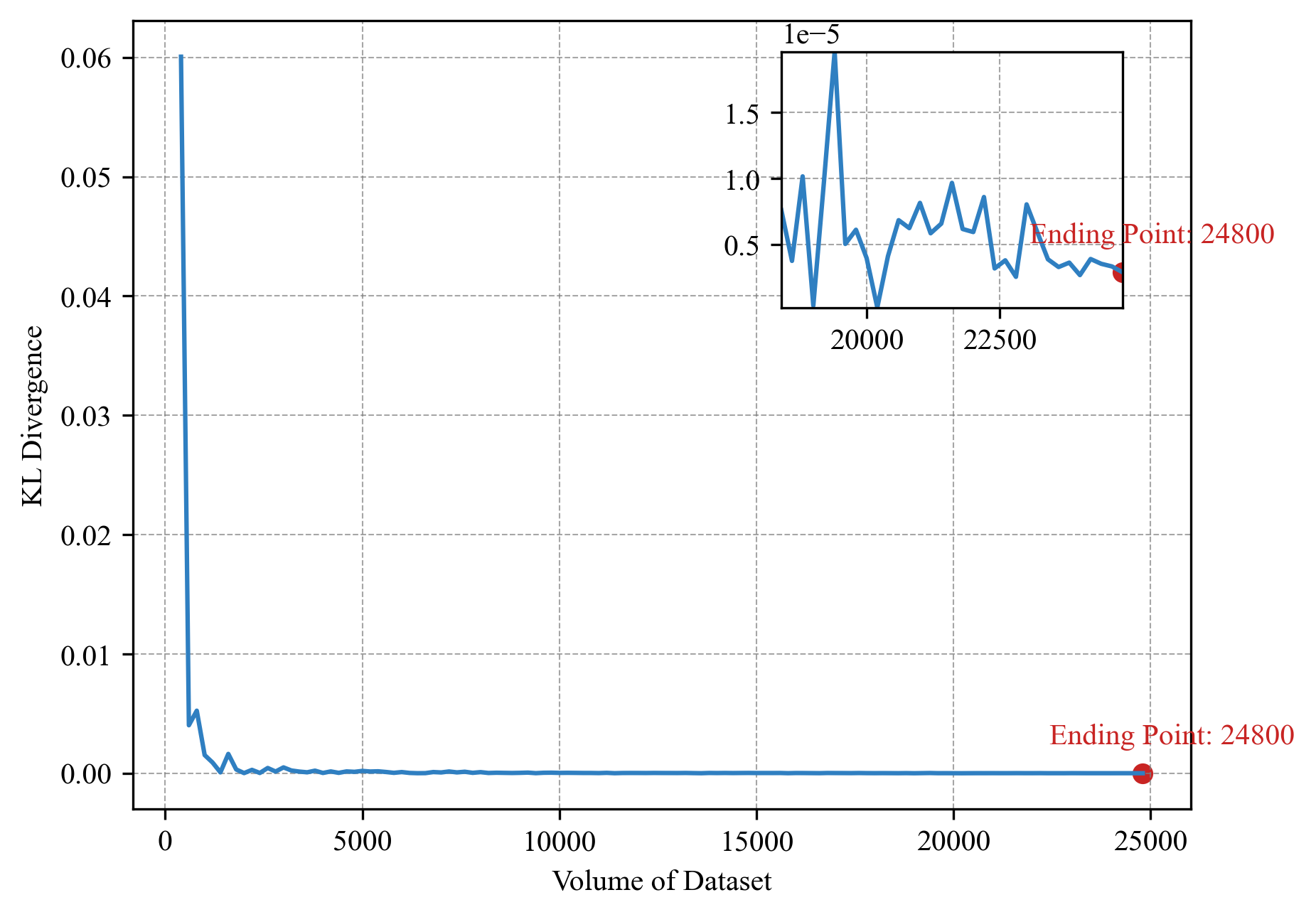}
        \caption{Balabit User12 Session7}
    \end{subfigure}
    \hfill
    \begin{subfigure}[b]{0.23\textwidth}
        \includegraphics[width=\textwidth]{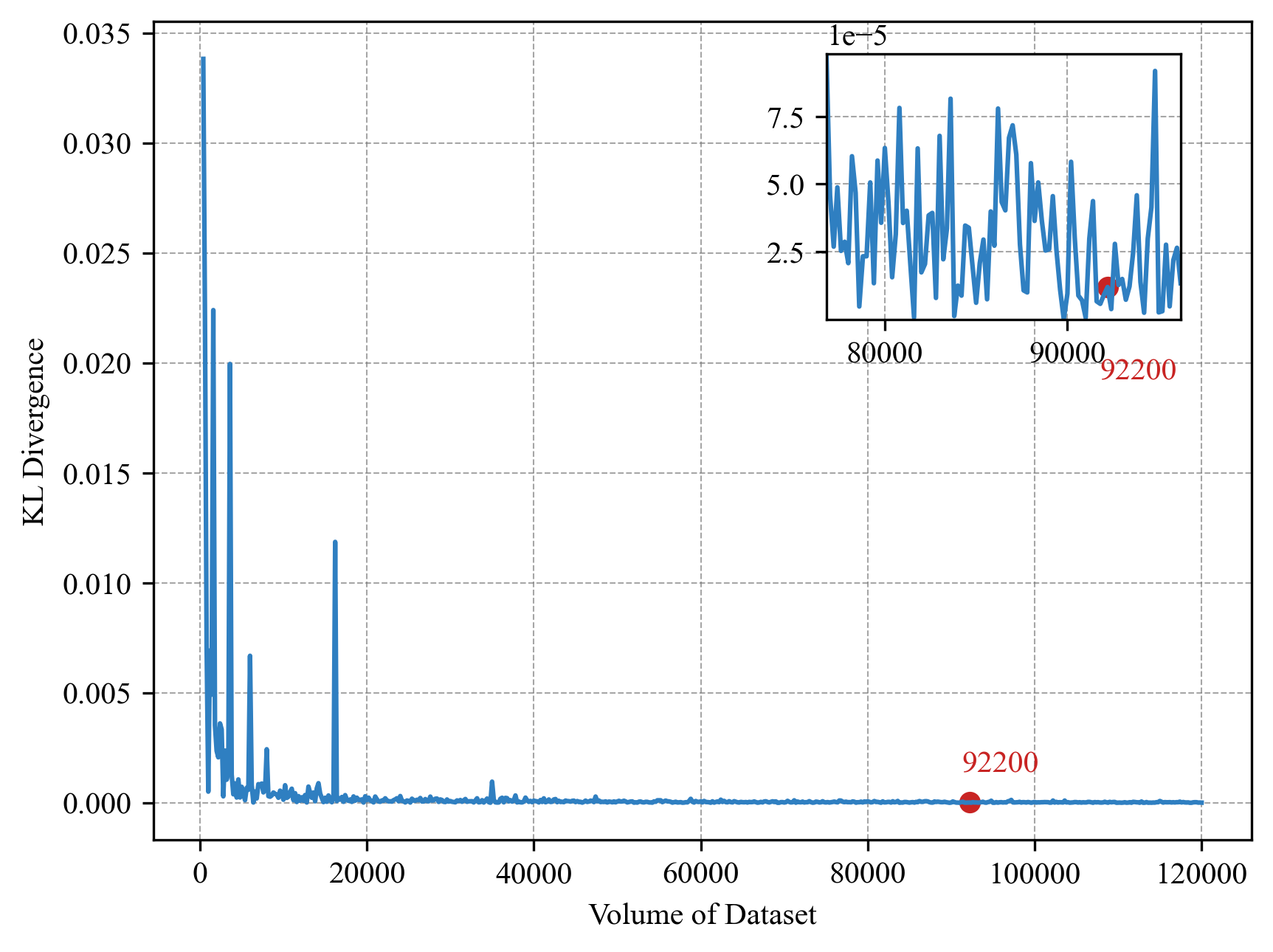}
        \caption{DFL User9}
    \end{subfigure}
    \caption{KL divergence convergence performance of User12 in the Balabit and User 9 in DFL datasets}
    \label{fig:user9_kl}
\end{figure*}

In the end, as shown in Figure \ref{fig:balabit_compare} and Figure \ref{fig:dflcompare}, the average suitable mouse dynamic dataset volume 73,778.7 of Balabit dataset is smaller than orginal one 114,114.2. For the DFL dataset, our method reduces the amount of data required for each user by a factor of ten, from $0.6726 \times 10^7$ to $0.0691 \times 10^7$.

\subsection{Mouse Authentication Unit Length and Model Performance Trade-off}  


\begin{figure*}[htbp]
    \centering
    \begin{subfigure}[b]{0.31\textwidth}
        \centering
        \includegraphics[width=\textwidth]{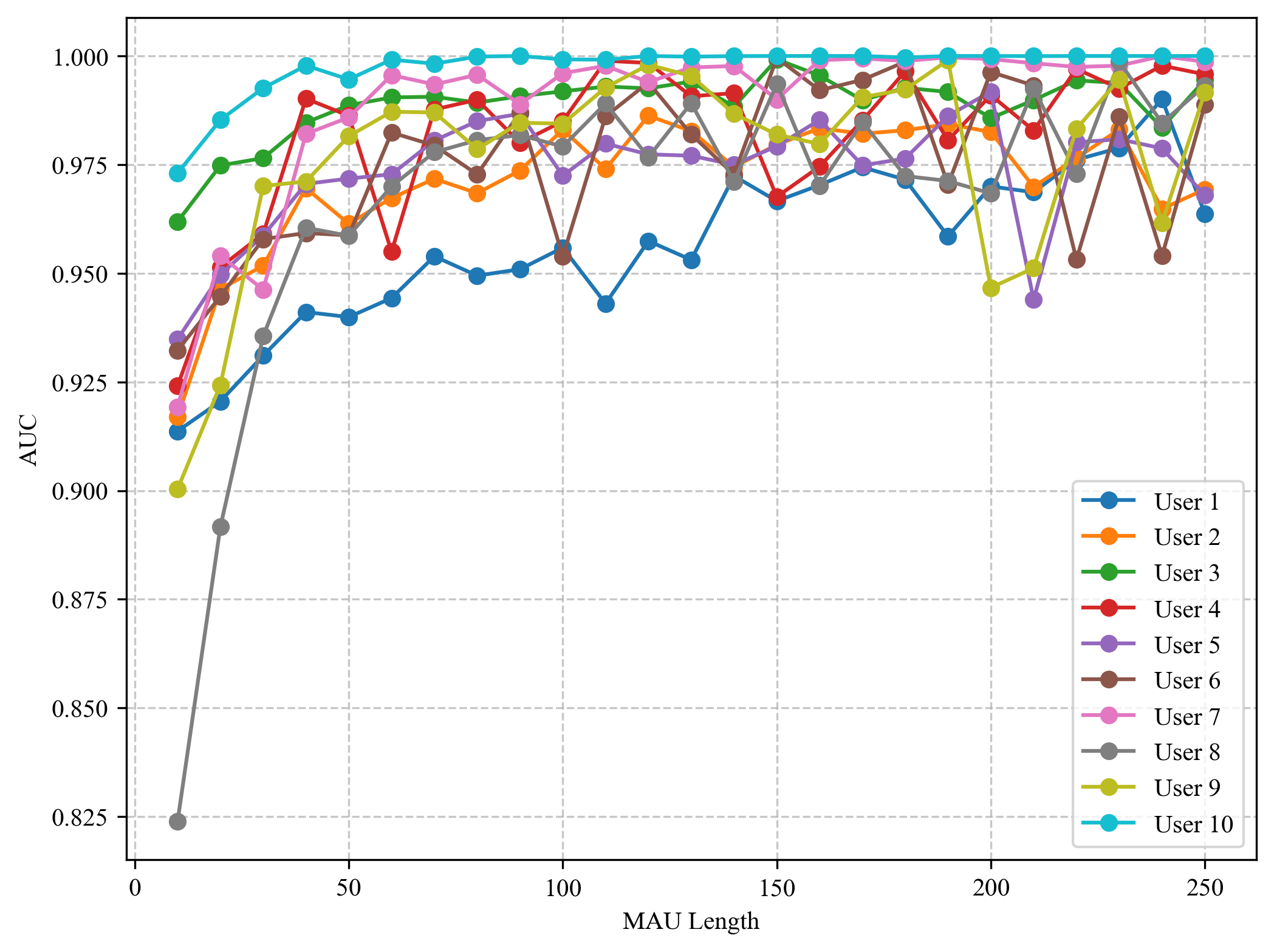}
        \caption{AUC Comparison}
        \label{fig:dfl_auc_1_10}
    \end{subfigure}
    \hfill
    \begin{subfigure}[b]{0.31\textwidth}
        \centering
        \includegraphics[width=\textwidth]{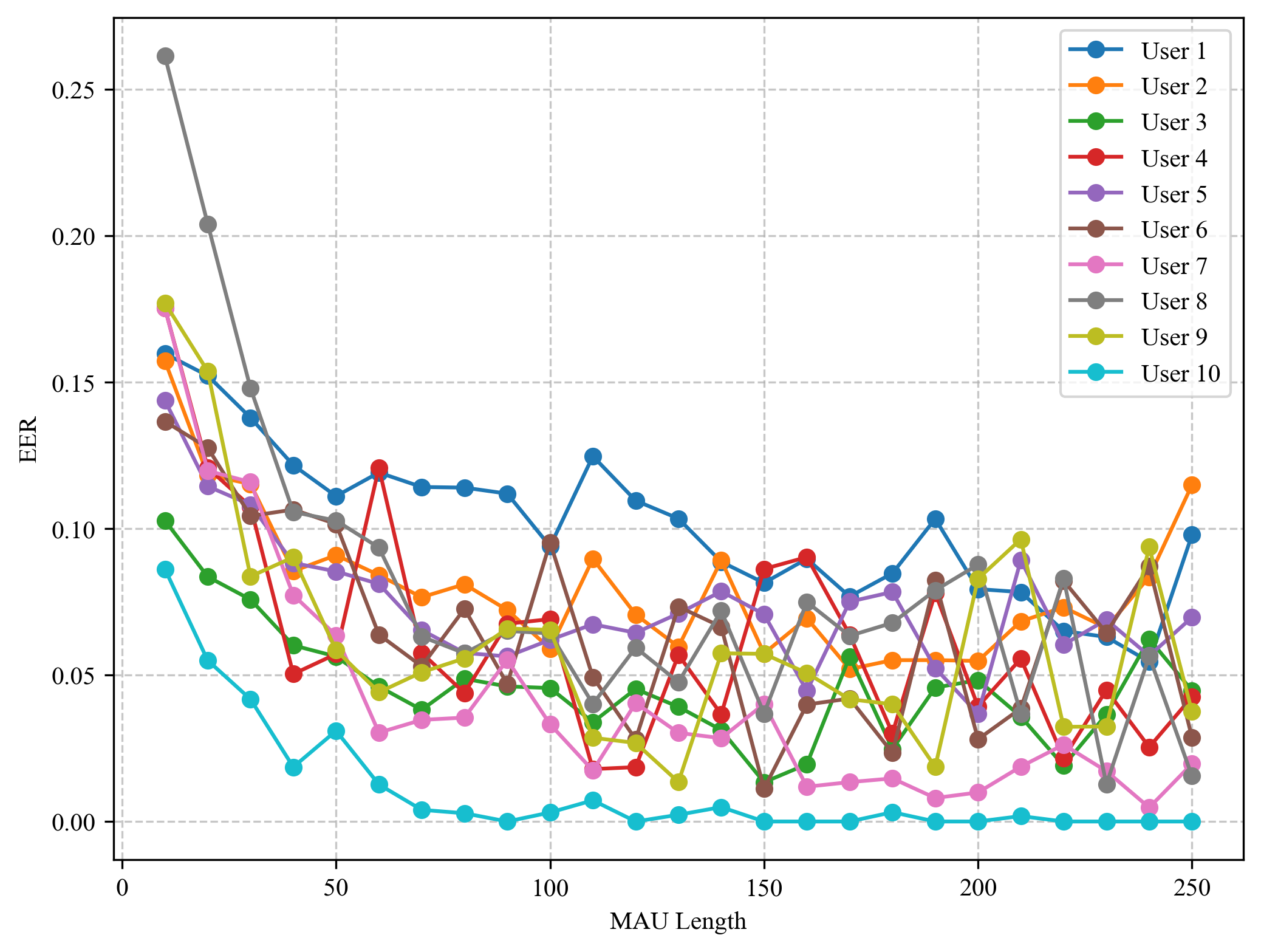}
        \caption{EER Comparison}
        \label{fig:dfl_eer_1_11}  
    \end{subfigure}
    \hfill
    \begin{subfigure}[b]{0.31\textwidth}
        \centering
        \includegraphics[width=\textwidth]{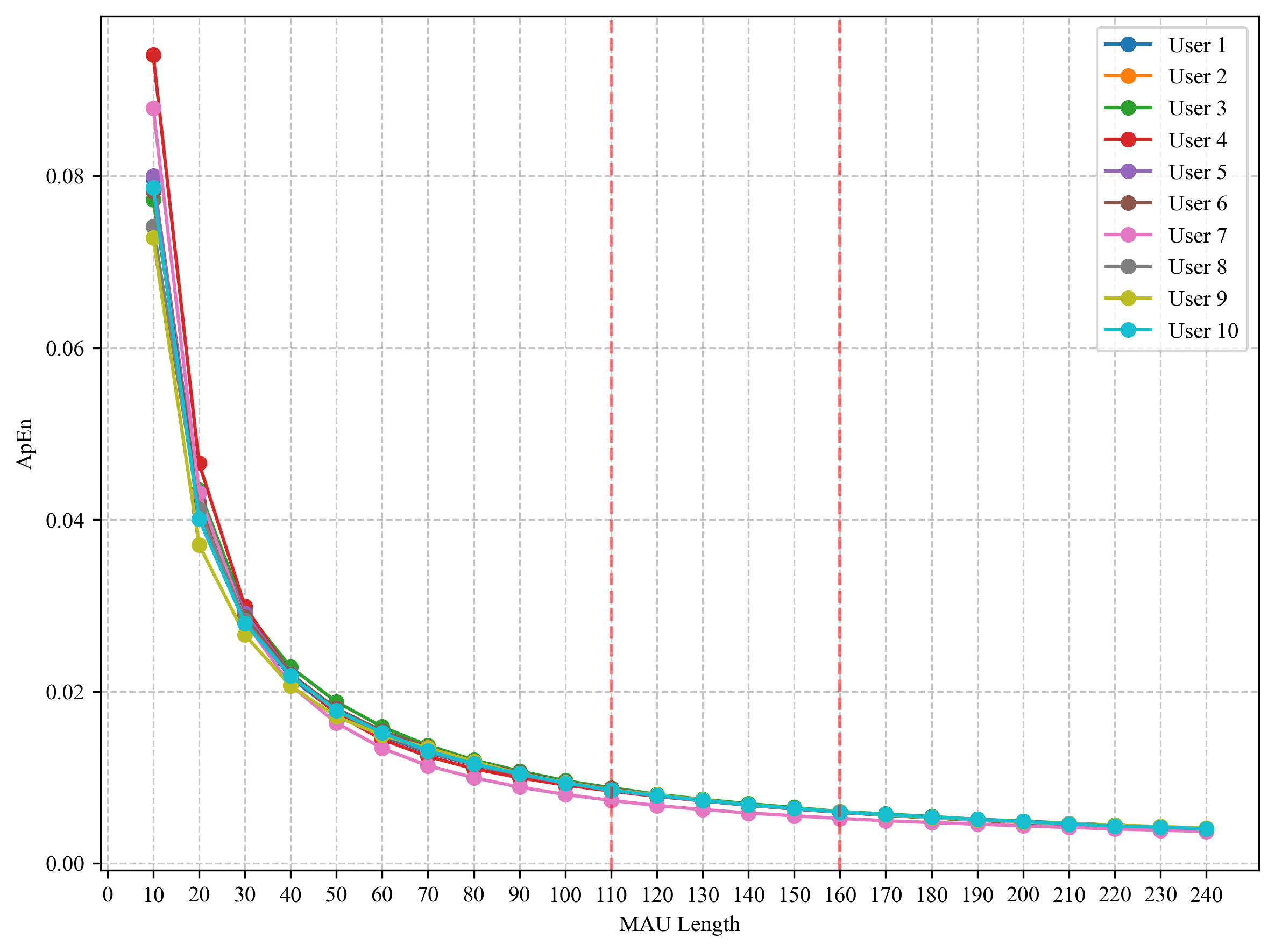}
        \caption{User Distribution}
        \label{fig:dfl_users_1_10}
    \end{subfigure}
    \caption*{DFL Dataset User 1-10}
    
    \vspace{1em}
    \begin{subfigure}[b]{0.31\textwidth}
        \centering
        \includegraphics[width=\textwidth]{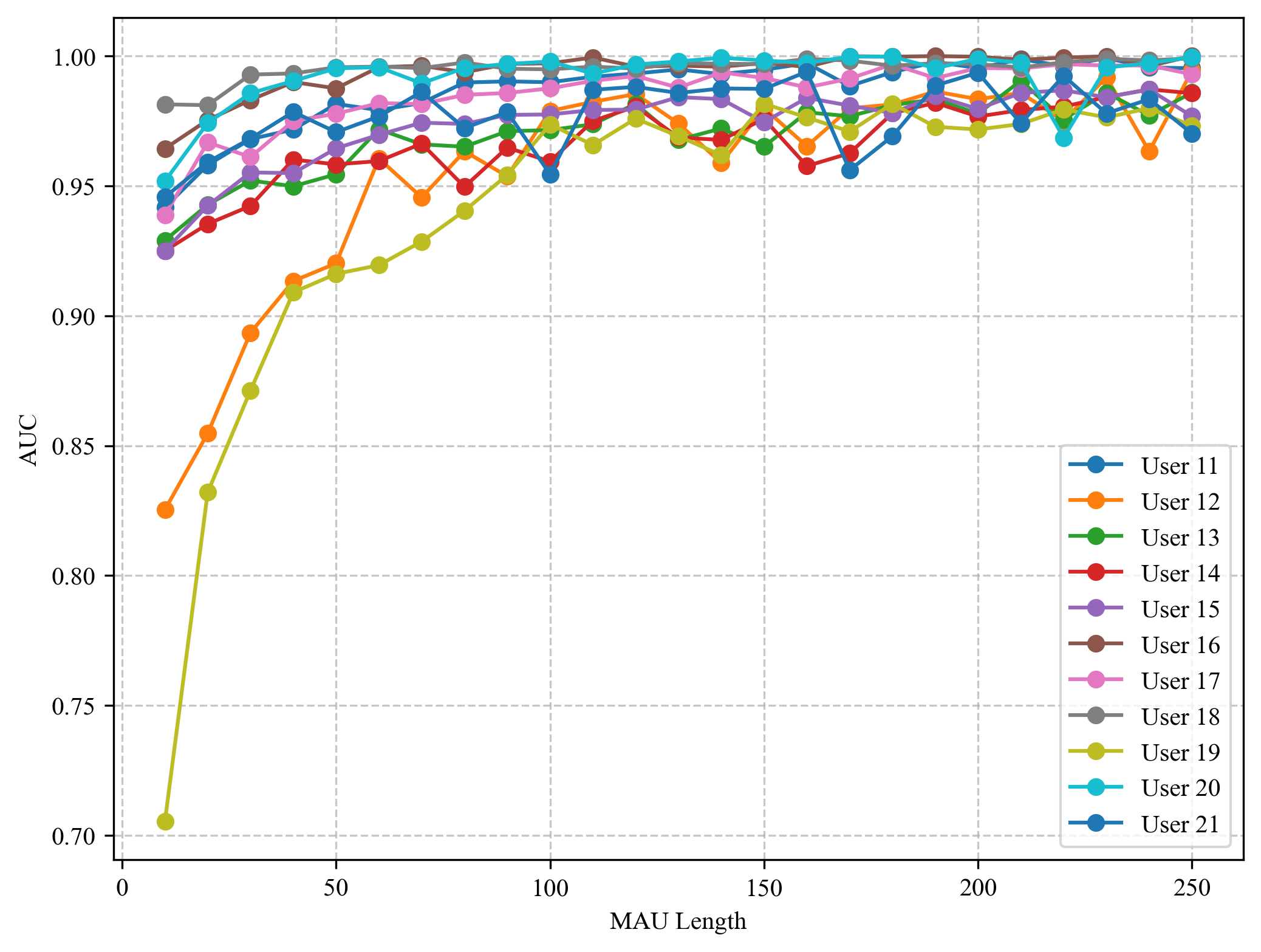}
        \caption{AUC Comparison}
        \label{fig:dfl_auc_11_21}
    \end{subfigure}
    \hfill
    \begin{subfigure}[b]{0.31\textwidth}
        \centering
        \includegraphics[width=\textwidth]{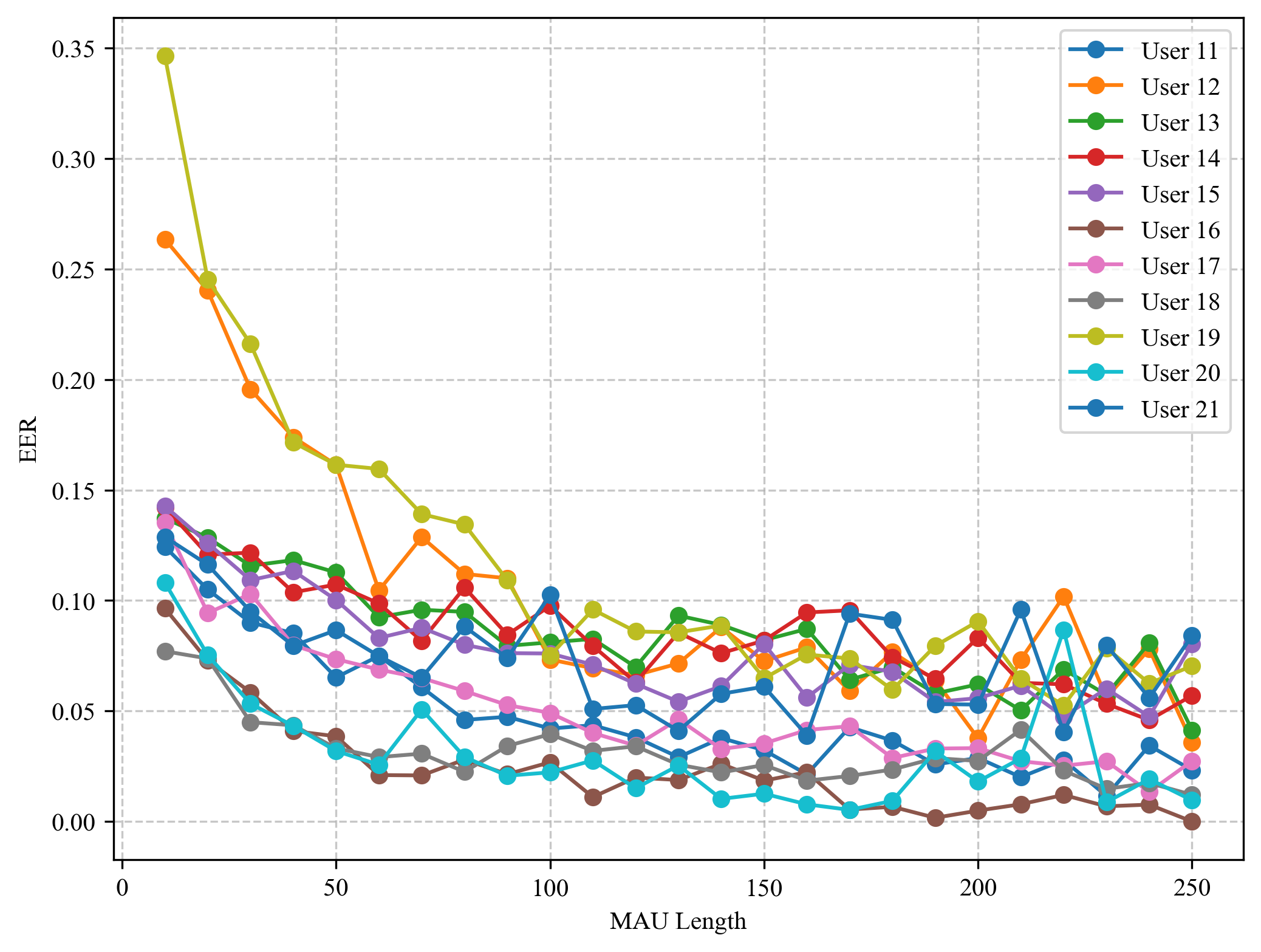}
        \caption{EER Comparison}
        \label{fig:dfl_eer_11_21}
    \end{subfigure}
    \hfill
    \begin{subfigure}[b]{0.31\textwidth}
        \centering
        \includegraphics[width=\textwidth]{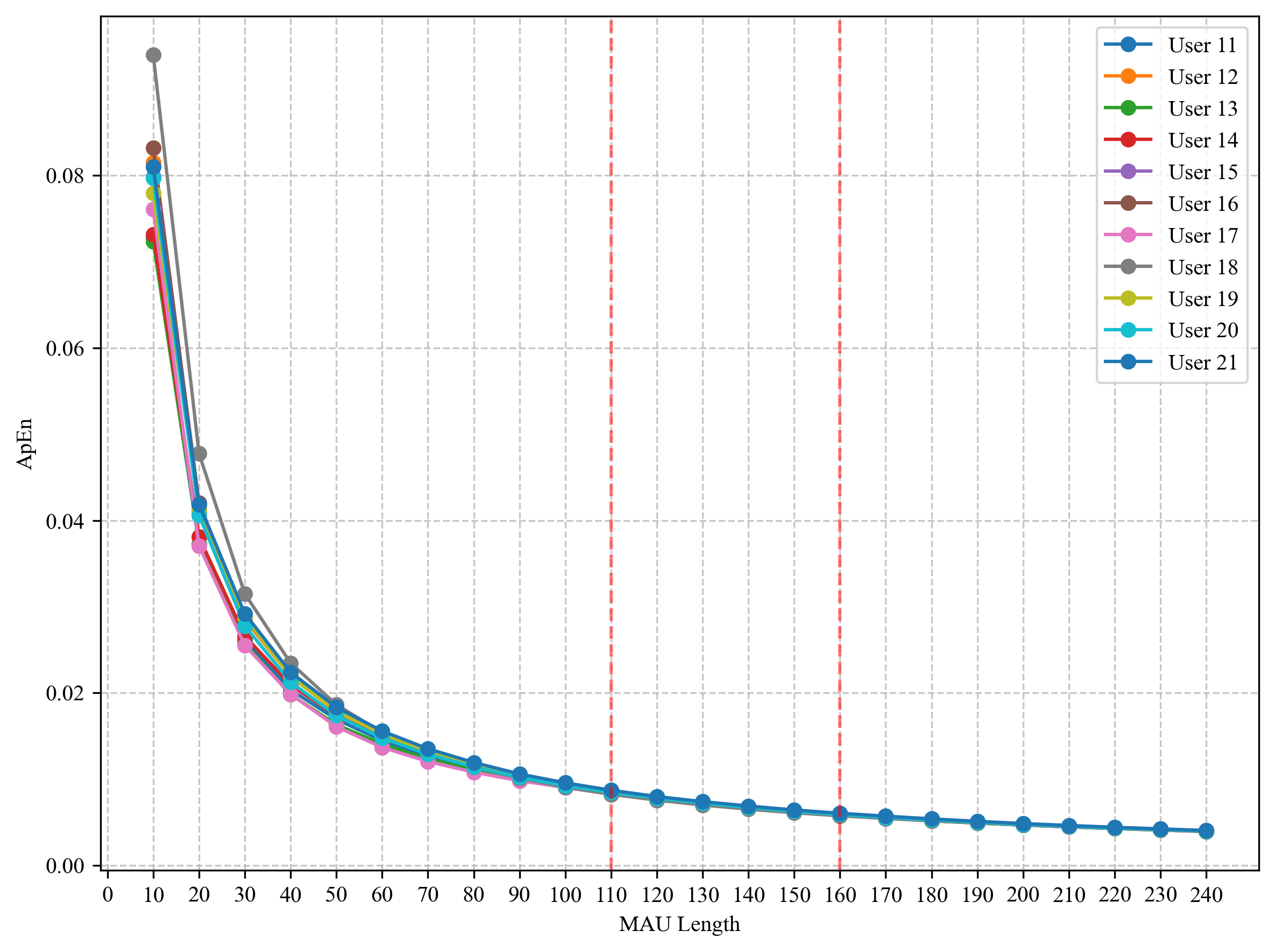}
        \caption{User Distribution}
        \label{fig:dfl_users_11_21}
    \end{subfigure}
    \caption*{DFL Dataset User 11-21}  
    
    \vspace{1em}
    \begin{subfigure}[b]{0.31\textwidth}
        \centering
        \includegraphics[width=\textwidth]{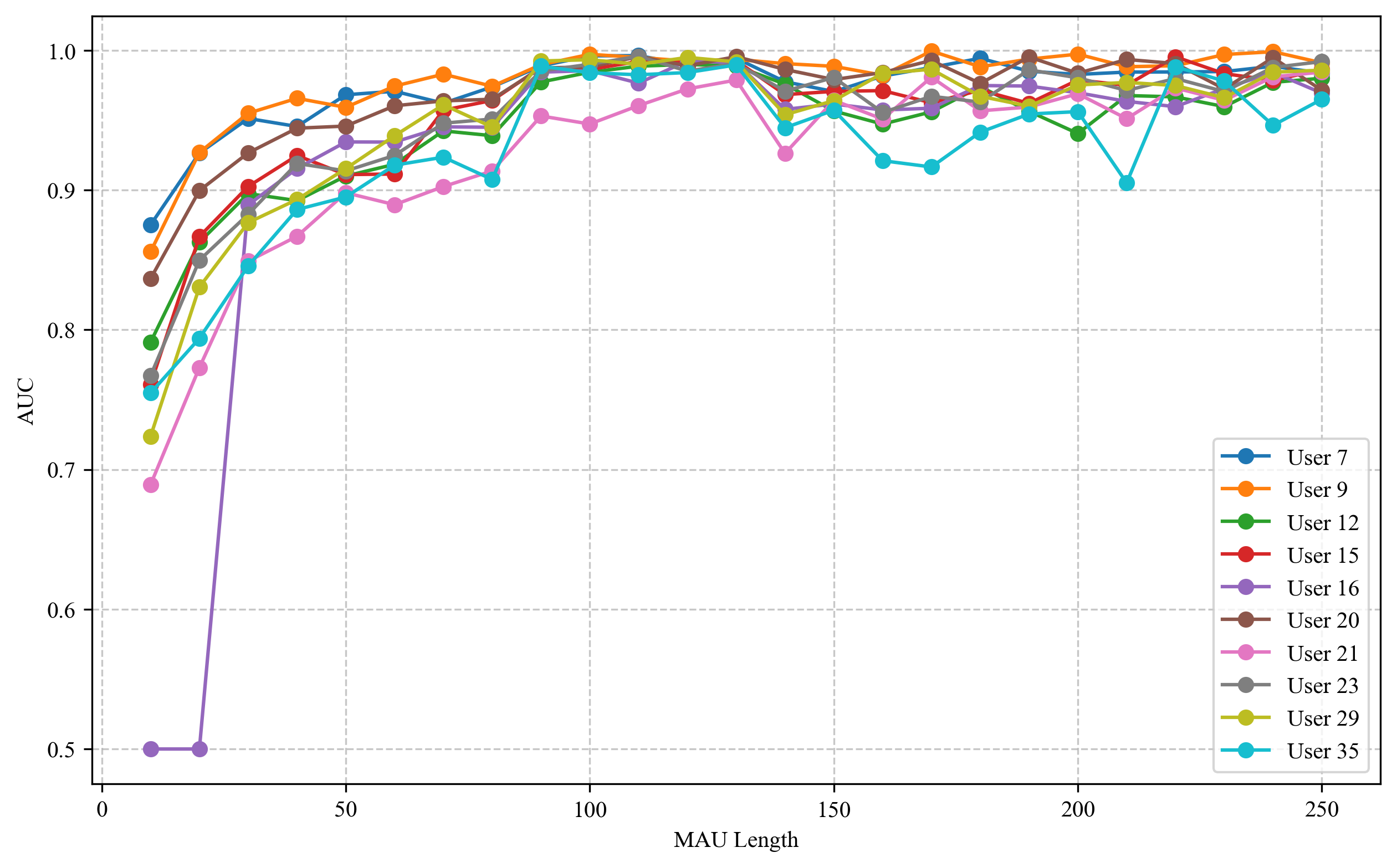}
        \caption{AUC Comparison}
        \label{fig:balabit_auc}
    \end{subfigure}
    \hfill
    \begin{subfigure}[b]{0.31\textwidth}
        \centering
        \includegraphics[width=\textwidth]{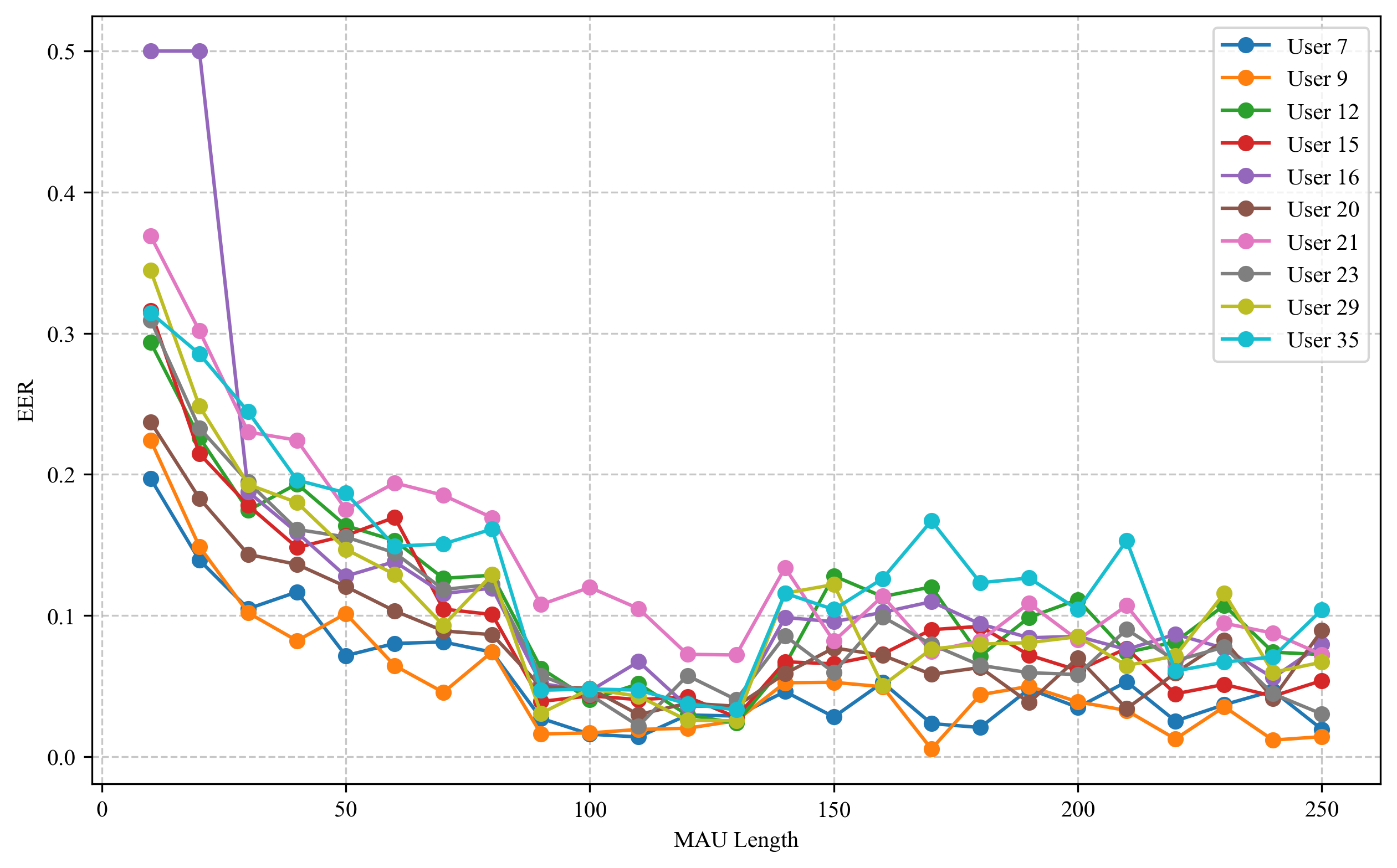}
        \caption{EER Comparison}
        \label{fig:balabit_eer}
    \end{subfigure}
    \hfill
    \begin{subfigure}[b]{0.31\textwidth}
        \centering
        \includegraphics[width=\textwidth]{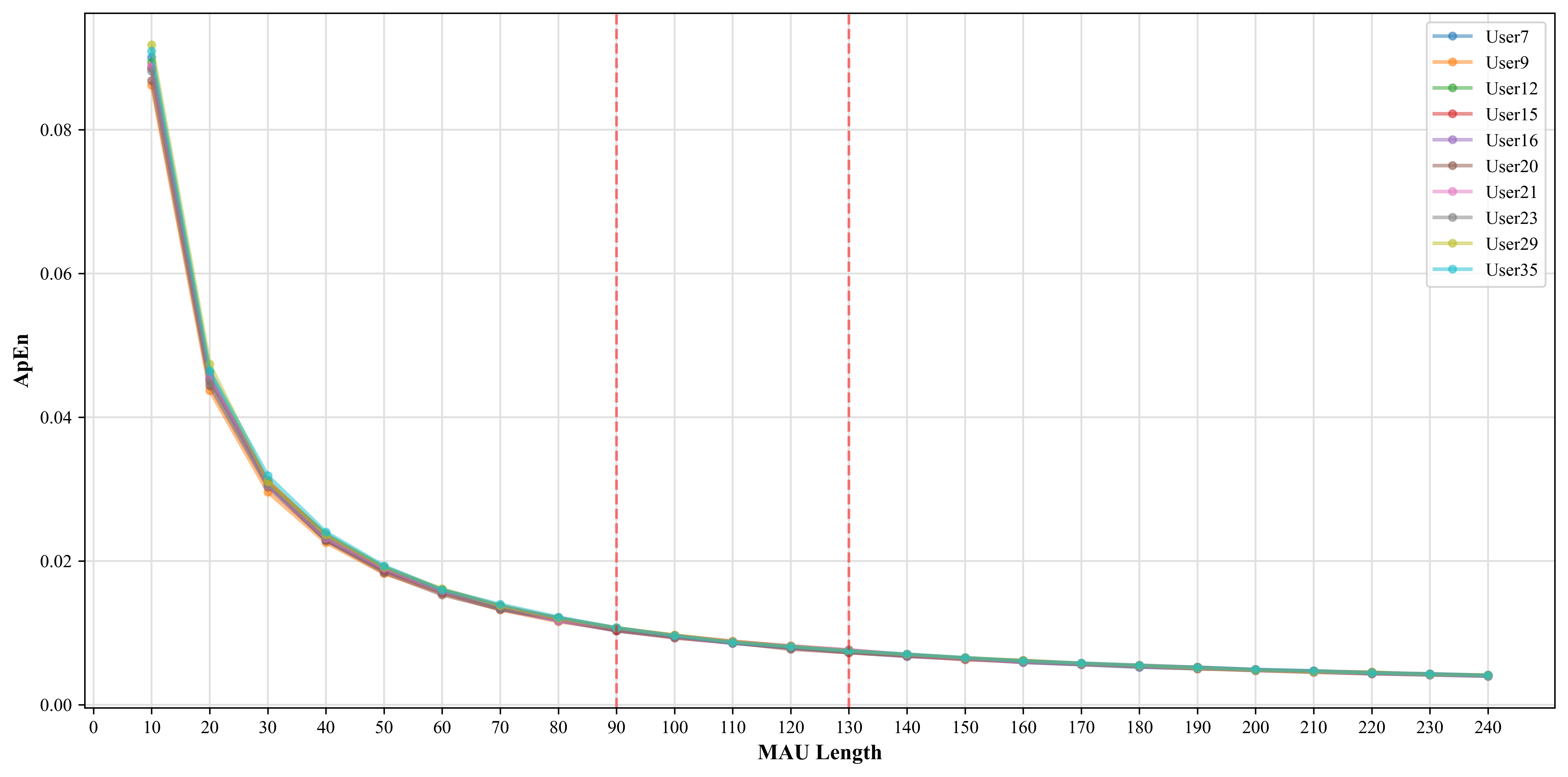}
        \caption{User Distribution}
        \label{fig:balabit_users}
    \end{subfigure}
    \caption*{Balabit Dataset}
    
    \caption{Performance Metrics and ApEn Analysis on DFL and Balabit Dataset}
    \label{fig:resultofall}  
\end{figure*}

In this section, We analyzed the performance impact of different MAU (Mouse Action Unit) lengths on the authentication model LT-AMouse in Balabit and DFL datasets. The results show that as the MAU length increases, the AUC index of the model gradually increases and the EER decreases accordingly. This is because longer MAUs provide richer mouse dynamics features, which help to portray user behavior more comprehensively. However, the performance of the model does not increase linearly. In the short MAU length range, the AUC and EER improve significantly, and when the length exceeds a certain threshold, the performance improvement tends to slow down.

Through the analysis of Approximate Entropy (ApEn), we find that this trend is consistent with the pattern of sequence randomness reduction. With a short MAU length, the data is not enough to fully characterize the user; as the length increases, the uncertainty of the sequence decreases, and the model can quickly accumulate discriminative features. However, when the data volume reaches a certain scale, the additional information gain decreases, and the change amplitude of both ApEn and performance indicators tends to level off.  

Therefore, the choice of MAU length needs to be a trade-off between security and real-time: high security requirement scenarios can appropriately increase the MAU length to improve accuracy, while applications with high real-time requirements need to shorten the MAU length to realize fast response. This study provides a reference basis for the practical deployment of mouse dynamics authentication system.

By further analyzing the slope of the change of entropy, we find that when the absolute value of the slope of the entropy is close to or less than $1\times10^{-4}$, the increase of the MAU length tends to moderate the enhancement of the model performance. At this point, the decrease in entropy is small, indicating that the accumulated feature information is close to saturation, and continuing to increase the MAU length has limited gain in model performance, while the computational cost may increase significantly. Therefore, an absolute value of the entropy slope less than $1\times10^{-4}$ is defined as the optimal balance between efficiency and model performance.  

It can be observed in Figure \ref{fig:resultofall} that the optimal equilibrium point for the Balabit dataset lies between 90-130 MAU lengths, while the DFL dataset lies between 110-160. This analysis provides a theoretical basis for the selection of experimental data in Section 7.4, helping us to optimize the efficiency while ensuring the model performance.  

\subsection{Comparison with Other Models}
To ensure a fair comparison with existing state-of-the-art methods, we evaluate our proposed user authentication model on the DFL and Balabit datasets under identical experimental settings. For consistency, we use the same hyper-parameter configurations, including the network architecture, loss function, and training epochs, as employed by baseline methods. This ensures that performance improvements are attributed solely to our model's design rather than differing experimental conditions.

Table~\ref{tab:model_comparison1} and table~\ref{tab:model_comparison2} summarizes the performance of our model compared to other commonly used methods, including CNN, LSTM, RF, and SVM. On the DFL dataset, our model achieves the highest F1 score (97.24\%) and AUC (98.52\%), while maintaining a low equal error rate (EER) of 5.05\%. Similarly, on the Balabit dataset, our model outperforms all baselines, achieving an F1 score of 94.65\%, AUC of 97.73\%, and an EER of 6.14\%. These results demonstrate that our model not only provides state-of-the-art accuracy but also ensures robustness across different datasets.  

Moreover, compared to traditional machine learning models such as RF and SVM, our model achieves significant improvements in both accuracy and efficiency. Specifically, our model is more than 8 times accurate than RF in terms of EER on the DFL dataset and reduces the error rate by over 40\% compared to SVM on the Balabit dataset. These enhancements highlight the strength of our approach in capturing the underlying dynamics of user-specific mouse behavior.  

Finally, our method exhibits superior generalizability across datasets, as evidenced by consistent improvements in both precision and recall. This capability is particularly critical for real-world applications, where datasets often vary significantly in terms of user behavior and interaction patterns.

\begin{table}[htbp]
\centering
\caption{User-Averaged Models Performance Comparison on DFL Dataset}
\label{tab:model_comparison1}
\begin{tabular}{cccc}
\toprule
\textbf{Model} & \textbf{F1} & \textbf{AUC} & \textbf{EER} \\
\midrule
Our Model & 97.24\% & 98.52\% & 5.05\% \\
          & (0.13\%) & (0.01\%) & (0.08\%) \\
\midrule
CNN       & 96.01\% & 97.07\% & 7.56\% \\
          & (0.68\%) & (0.07\%) & (0.18\%) \\
\midrule
LSTM      & 94.22\% & 83.48\% & 23.11\% \\
          & (0.07\%) & (1.66\%) & (1.45\%) \\
\midrule
RF        & 79.53\% & 89.85\% & 14.74\% \\
          & (2.56\%) & (1.43\%) & (1.71\%) \\
\midrule
SVM       & 88.92\% & 59.40\% & 42.97\% \\
          & (0.00\%) & (2.77\%) & (2.21\%) \\
\bottomrule
\end{tabular}
\end{table}

\begin{table}[htbp]
\centering
\caption{User-Averaged Models Performance Comparison on Balabit Dataset}
\label{tab:model_comparison2}  
\begin{tabular}{cccc}
\toprule
\textbf{Model} & \textbf{F1} & \textbf{AUC} & \textbf{EER} \\
\midrule
\multirow{2}{*}{\centering Our Model} & 94.65$\%$ & 97.73$\%$ & 6.14$\%$ \\
& (0.71$\%$) & (0.03$\%$) & (0.11$\%$) \\
\midrule
\multirow{2}{*}{\centering CNN} & 93.01$\%$ & 93.15$\%$ & 13.79$\%$ \\
& (0.28$\%$) & (0.09$\%$) & (0.20$\%$) \\
\midrule
\multirow{2}{*}{\centering LSTM} & 89.79$\%$ & 80.18$\%$ & 25.85$\%$ \\  
& (0.24$\%$) & (1.35$\%$) & (1.21$\%$) \\
\midrule
\multirow{2}{*}{\centering RF} & 54.62$\%$ & 72.18$\%$ & 33.36$\%$ \\
& (0.43$\%$) & (1.63$\%$) & (1.21$\%$) \\
\midrule
\multirow{2}{*}{\centering SVM} & 86.10$\%$ & 44.08$\%$ & 54.04$\%$ \\
& (0.07$\%$) & (2.47$\%$) & (1.48$\%$) \\
\bottomrule
\end{tabular}
\end{table}

\subsection{Attack Model}
We assume that the attacker is familiar with the authentication mechanism of LT-AMouse. Depending on whether the adversary can access the parameters of the LT-AMouse model and whether they can obtain partial mouse movement data from legitimate users (e.g., through phishing emails or other malicious means), we classify attacks into the following two types:      

\textbf{Blind Attack} In a blind attack, the adversary possesses no prior knowledge of the legitimate user's mouse movement patterns. To carry out the attack, the adversary interacts with the system by controlling the mouse, attempting to bypass the authentication mechanism using their own mouse dynamics.  

\textbf{\textit{Result}} Blind attack results are summarized in Table \ref{tab:model_comparison1} and Table \ref{tab:model_comparison2}. By introducing unseen samples from the LT-AMouse dataset into the test set, which were not present in the training set, our model achieved notable performance across all evaluation metrics. Specifically, on , the model achieved an F1 Score of 97.24

\textbf{Imitation Attack} In an imitation attack, we assume that the adversary can observe, record, and analyze the dynamic mouse trajectories of legitimate users. Leveraging this data, the adversary employs generative models\cite{Tan2019AdversarialAttacks} to create highly realistic forged mouse trajectories, aiming to deceive the authentication system and impersonate a legitimate user.  

In this study, we employ a tailored attack model designed to circumvent a mouse dynamics authentication system. This model leverages the Wasserstein Conditional Deep Convolutional Generative Adversarial Network (WCDCGAN) \cite{Roy2023WCDCGAN} to generate realistic and high-quality adversarial samples that closely mimic genuine mouse dynamics, making them challenging for the authentication system to distinguish from legitimate inputs.

\textbf{\textit{Result}}
In the simulation attack scenario, this study employs DSR as the core evaluation metric to assess LT-AMouse's capability in distinguishing between legitimate users and generated fraudulent data under different parameter configurations. Table \ref{tab:attackmodel} demonstrates the system's defensive performance against imitation attacks when the MAU length increases from 110 to 160.

\begin{table}[htbp]  
    \centering
    \caption{Defense Success Rates for Different MAU Lengths}
    \label{tab:defense_rates}
    \begin{tabular}{lcc}
    \toprule
    Dataset & MAU Length & Defense Success Rate (\%) \\ 
    \midrule
    \multirow{6}{*}{DFL} & 110 & 89.22 \\
    & 120 & 99.02 \\
    & 130 & 88.56 \\
    & 140 & 95.87 \\
    & 150 & 94.44 \\
    & 160 & 93.68 \\
    \midrule
    \multirow{5}{*}{Balabit} & 90 & 47.51 \\
    & 100 & 70.00 \\
    & 110 & 79.69 \\
    & 120 & 59.90 \\
    & 130 & 60.00 \\
    \bottomrule
    \end{tabular}
    \label{tab:attackmodel}
\end{table}

As shown in Table \ref{tab:attackmodel}, for the DFL dataset, when the MAU length is 120, the system achieves a defense success rate of up to 99.02\%. At other lengths the defense success rate remains between 88.56\% and 95.87\%. For the Balabit dataset, our model achieves the best defense success rate of 79.69\% at MAU length 110, which demonstrates a significant improvement compared to the original attack success rate of 94\% reported in the WC-DCGAN attack \cite{Roy2023WCDCGAN} . These findings suggest that the proposed LT-AMouse model can accurately differentiate genuine from forged mouse trajectories, even with high-fidelity imitation data, thereby demonstrating strong robustness and security.

\section{Discussion}
\textbf{Environmental interference and minimal user input} In practical applications, mouse dynamics-based authentication relies on the consistent capture of movement trajectories and click patterns. However, significant fluctuations in surface friction or hardware sensitivity caused by environmental factors, as well as minimal and low-amplitude mouse operations, pose challenges to accurate modeling of these sparse and weak dynamic signals, leading to performance degradation. Compared to continuous and pronounced mouse inputs, unstable or low-amplitude operations are more susceptible to environmental factors and usage posture, increasing the likelihood of errors in authentication.   

\textbf{Biometric Authentication} Mouse operation habits are susceptible to variations in hardware, environment, and users' states (e.g. fatigue and emotions), leading to significant shifts in data distribution over time and weak generalization. Additionally, in cross-platform or cross-context deployments, mouse movement features can vary substantially due to differences in operating systems or the mouse's settings, posing challenges to model transferability. Furthermore, while mouse-dynamics-based approaches offer advantages in data collection and privacy protection compared to biometic methods, they remain vulnerable to some threats. For example, attackers could potentially infer user behavior patterns and compromise system security by intercepting portions of mouse trajectories through malicious software or remote monitoring. Lastly, to continuously enhance security and robustness, further research is needed in privacy-protection technologies, such as on-device computation, federated learning, and secure multiparty computation, along with in-depth model adaptation and optimization tailored to diverse application scenarios.  

\textbf{Attack Model} In this study, we assessed our method’s resilience against the advanced WCDCGAN \cite{Roy2023WCDCGAN} attack, which has demonstrated up to a 94\% success rate in compromising mouse dynamics authentication systems. Notably, our approach can operate effectively with smaller datasets, thus enabling faster and more efficient defense performance—albeit without achieving complete immunity. Future research will focus on integrating advanced strategies and novel techniques to further fortify the system’s resistance. Empirical results show that our method achieved defense success rates exceeding 88\% on the DFL dataset, whereas the Balabit dataset peaked at 79.69\%, suggesting that dataset properties significantly influence how well the system can withstand sophisticated adversarial threats. 

\section{Conclusion}

This study presents a robust and efficient mouse dynamics-based authentication framework, addressing challenges in data sufficiency, practicality, and security. By introducing the Mouse Authentication Unit (MAU) and leveraging Approximate Entropy, our method optimizes data segmentation for accurate behavioral representation. The Local-Time Mouse Authentication (LT-MAuthen) framework achieved state-of-the-art performance, with AUCs of $98.52\%$ on the DFL dataset and $94.65\%$ on the Balabit dataset, while demonstrating resilience against advanced adversarial attacks. These findings highlight the framework's potential for real-world applications, with future work focused on enhancing cross-platform adaptability and robust defense mechanisms.


\begin{thebibliography}{10}
\providecommand{\url}[1]{#1}
\csname url@samestyle\endcsname
\providecommand{\newblock}{\relax}
\providecommand{\bibinfo}[2]{#2}
\providecommand{\BIBentrySTDinterwordspacing}{\spaceskip=0pt\relax}
\providecommand{\BIBentryALTinterwordstretchfactor}{4}
\providecommand{\BIBentryALTinterwordspacing}{\spaceskip=\fontdimen2\font plus
\BIBentryALTinterwordstretchfactor\fontdimen3\font minus \fontdimen4\font\relax}
\providecommand{\BIBforeignlanguage}[2]{{%
\expandafter\ifx\csname l@#1\endcsname\relax
\typeout{** WARNING: IEEEtran.bst: No hyphenation pattern has been}%
\typeout{** loaded for the language `#1'. Using the pattern for}%
\typeout{** the default language instead.}%
\else
\language=\csname l@#1\endcsname
\fi
#2}}
\providecommand{\BIBdecl}{\relax}
\BIBdecl

\bibitem{erlich2009authentication}
Z.~Erlich and M.~Zviran, ``Authentication methods for computer systems security,'' in \emph{Encyclopedia of Information Science and Technology, Second Edition}.\hskip 1em plus 0.5em minus 0.4em\relax IGI Global, 2009, pp. 288--293.

\bibitem{proctor2002improving}
R.~W. Proctor, M.-C. Lien, K.-P.~L. Vu, E.~E. Schultz, and G.~Salvendy, ``Improving computer security for authentication of users: Influence of proactive password restrictions,'' \emph{Behavior Research Methods, Instruments, \& Computers}, vol.~34, pp. 163--169, 2002.

\bibitem{revett2008survey}
K.~Revett, H.~Jahankhani, S.~T.~D. Magalhaes, and H.~M.~D. Santos, ``A survey of user authentication based on mouse dynamics,'' in \emph{Global E-Security: 4th International Conference, ICGeS 2008, London, UK, June 23-25, 2008. Proceedings}.\hskip 1em plus 0.5em minus 0.4em\relax Springer, 2008, pp. 210--219.

\bibitem{haussler1997general}
D.~Haussler, ``A general minimax result for relative entropy,'' \emph{IEEE Transactions on Information Theory}, vol.~43, no.~4, pp. 1276--1280, 1997.

\bibitem{villani2000short}
C.~Villani, ``A short proof of the "concavity of entropy power",'' \emph{IEEE Transactions on Information Theory}, vol.~46, no.~4, pp. 1695--1696, 2000.

\bibitem{he2016deep}
K.~He, X.~Zhang, S.~Ren, and J.~Sun, ``Deep residual learning for image recognition,'' in \emph{Proceedings of the IEEE Conference on Computer Vision and Pattern Recognition}, 2016, pp. 770--778.

\bibitem{kingma2014adam}
D.~P. Kingma and J.~Ba, ``Adam: A method for stochastic optimization,'' \emph{arXiv preprint arXiv:1412.6980}, 2014.

\bibitem{wang2021hybrid}
K.~Wang, C.~Ma, Y.~Qiao, X.~Lu, W.~Hao, and S.~Dong, ``A hybrid deep learning model with 1dcnn-lstm-attention networks for short-term traffic flow prediction,'' \emph{Physica A: Statistical Mechanics and its Applications}, vol. 583, p. 126293, 2021.

\bibitem{fu2020rumba}
S.~Fu, D.~Qin, D.~Qiao, and G.~T. Amariucai, ``Rumba-mouse: Rapid user mouse-behavior authentication using a cnn-rnn approach,'' in \emph{2020 IEEE Conference on Communications and Network Security (CNS)}.\hskip 1em plus 0.5em minus 0.4em\relax IEEE, 2020, pp. 1--9.

\bibitem{chong2019user}
P.~Chong, Y.~Elovici, and A.~Binder, ``User authentication based on mouse dynamics using deep neural networks: A comprehensive study,'' \emph{IEEE Transactions on Information Forensics and Security}, vol.~15, pp. 1086--1101, 2019.

\bibitem{yao2020identity}
Q.~Yao, J.~Zhao, Z.~Yang, R.~Fei, L.~Yan, and Y.~Wang, ``Identity authentication based on user mouse behavior,'' in \emph{2020 International Conference on Virtual Reality and Intelligent Systems (ICVRIS)}.\hskip 1em plus 0.5em minus 0.4em\relax IEEE, 2020, pp. 571--577.

\bibitem{antal2021sapimouse}
M.~Antal, N.~Fejér, and K.~Buza, ``Sapimouse: Mouse dynamics-based user authentication using deep feature learning,'' in \emph{2021 IEEE 15th International Symposium on Applied Computational Intelligence and Informatics (SACI)}.\hskip 1em plus 0.5em minus 0.4em\relax IEEE, 2021, pp. 61--66.

\bibitem{polemi1997biometric}
D.~Polemi, ``Biometric techniques: Review and evaluation of biometric techniques for identification and authentication, including an appraisal of the areas where they are most applicable,'' \emph{Reported Prepared for the European Commission DG XIIIC}, vol.~4, 1997.

\bibitem{joyce1990identity}
R.~Joyce and G.~Gupta, ``Identity authentication based on keystroke latencies,'' \emph{Communications of the ACM}, vol.~33, no.~2, pp. 168--176, 1990.

\bibitem{bailey2014user}
K.~O. Bailey, J.~S. Okolica, and G.~L. Peterson, ``User identification and authentication using multi-modal behavioral biometrics,'' \emph{Computers \& Security}, vol.~43, pp. 77--89, 2014.

\bibitem{meng2013touch}
Y.~Meng, D.~S. Wong, R.~Schlegel, and L.~for Kwok, ``Touch gestures based biometric authentication scheme for touchscreen mobile phones,'' in \emph{Information Security and Cryptology: 8th International Conference, Inscrypt 2012, Beijing, China, November 28-30, 2012, Revised Selected Papers 8}.\hskip 1em plus 0.5em minus 0.4em\relax Springer, 2013, pp. 331--350.

\bibitem{buriro2015touchstroke}
A.~Buriro, B.~Crispo, F.~D. Frari, and K.~Wrona, ``Touchstroke: Smartphone user authentication based on touch-typing biometrics,'' in \emph{New Trends in Image Analysis and Processing--ICIAP 2015 Workshops: ICIAP 2015 International Workshops, BioFor, CTMR, RHEUMA, ISCA, MADiMa, SBMI, and QoEM, Genoa, Italy, September 7-8, 2015, Proceedings 18}.\hskip 1em plus 0.5em minus 0.4em\relax Springer, 2015, pp. 27--34.

\bibitem{amarasinghe2021stress}
A.~M. Amarasinghe, I.~Malassri, K.~Weerasinghe, I.~Jayasingha, P.~K. Abeygunawardhana, and S.~Silva, ``Stress analysis and care prediction system for online workers,'' in \emph{2021 3rd International Conference on Advancements in Computing (ICAC)}.\hskip 1em plus 0.5em minus 0.4em\relax IEEE, 2021, pp. 329--334.

\bibitem{yi44trustworthy}
Q.~Yi, W.~Li, S.~ping Yi, and J.~dong Xie, ``Trustworthy identity authentication based on joint time-frequency analysis of mouse behavior,'' \emph{Journal of Beijing University of Posts and Telecommunications}, vol.~44, no.~4, p. 121.

\bibitem{ding2019user}
X.~Ding, C.~Peng, H.~Ding, M.~Wang, H.~Yang, and Q.~Yu, ``User identity authentication and identification based on multi-factor behavior features,'' in \emph{2019 IEEE Globecom Workshops (GC Wkshps)}.\hskip 1em plus 0.5em minus 0.4em\relax IEEE, 2019, pp. 1--6.

\bibitem{quraishi2022secure}
S.~J. Quraishi and S.~Bedi, ``Secure system of continuous user authentication using mouse dynamics,'' in \emph{2022 3rd International Conference on Intelligent Engineering and Management (ICIEM)}.\hskip 1em plus 0.5em minus 0.4em\relax IEEE, 2022, pp. 138--144.

\bibitem{siddiqui2021continuous}
N.~Siddiqui, R.~Dave, and N.~Seliya, ``Continuous user authentication using mouse dynamics, machine learning, and minecraft,'' in \emph{2021 International Conference on Electrical, Computer and Energy Technologies (ICECET)}.\hskip 1em plus 0.5em minus 0.4em\relax IEEE, 2021, pp. 1--6.

\bibitem{shen2010user}
C.~Shen, Z.~Cai, X.~Guan, C.~Fang, and Y.~Du, ``User authentication and monitoring based on mouse behavioural features,'' \emph{Journal on Communications}, vol.~31, no.~7, pp. 68--75, 2010.

\bibitem{ahmed2007new}
A.~A.~E. Ahmed and I.~Traore, ``A new biometric technology based on mouse dynamics,'' \emph{IEEE Transactions on Dependable and Secure Computing}, vol.~4, no.~3, pp. 165--179, 2007.

\bibitem{jun2019three}
H.~Jun and M.~Kang, ``Three-way identity authentication method based on mouse behavior,'' \emph{Journal of Nanjing University of Science and Technology}, no.~4, pp. 474--480, 2019.

\bibitem{antal2020mouse}
M.~Antal and N.~Fejér, ``Mouse dynamics based user recognition using deep learning,'' \emph{Acta Universitatis Sapientiae, Informatica}, vol.~12, no.~1, pp. 39--50, 2020.

\bibitem{komogortsev2015attack}
O.~V. Komogortsev, A.~Karpov, and C.~D. Holland, ``Attack of mechanical replicas: Liveness detection with eye movements,'' \emph{IEEE Transactions on Information Forensics and Security}, vol.~10, no.~4, pp. 716--725, 2015.

\bibitem{dai2016r}
J.~Dai, Y.~Li, K.~He, and J.~Sun, ``R-fcn: Object detection via region-based fully convolutional networks,'' \emph{Advances in Neural Information Processing Systems}, vol.~29, 2016.

\bibitem{li2016pruning}
H.~Li, A.~Kadav, I.~Durdanovic, H.~Samet, and H.~P. Graf, ``Pruning filters for efficient convnets,'' \emph{arXiv Preprint arXiv:1608.08710}, 2016.

\bibitem{huang2021makes}
Y.~Huang, C.~Du, Z.~Xue, X.~Chen, H.~Zhao, and L.~Huang, ``What makes multi-modal learning better than single (provably),'' \emph{Advances in Neural Information Processing Systems}, vol.~34, pp. 10\,944--10\,956, 2021.

\bibitem{pincus1995approximate}
S.~Pincus, ``Approximate entropy (apen) as a complexity measure,'' \emph{Chaos: An Interdisciplinary Journal of Nonlinear Science}, vol.~5, no.~1, pp. 110--117, 1995.

\bibitem{chiu2018mathematical}
E.~Chiu, J.~Lin, B.~McFerron, N.~Petigara, and S.~Seshasai, ``Mathematical theory of claude shannon. a study of the style and context of his work up to the genesis of information theory,'' \emph{Submitted for The Structure of Engineering Revolutions (MIT course 6.933 J/STS. 420J), nd}, 2018.

\bibitem{paysarvi2010optimal}
P.~Paysarvi-Hoseini and N.~C. Beaulieu, ``Optimal wideband spectrum sensing framework for cognitive radio systems,'' \emph{IEEE Transactions on Signal Processing}, vol.~59, no.~3, pp. 1170--1182, 2010.

\bibitem{fredkin2003introduction}
E.~Fredkin, ``An introduction to digital philosophy,'' \emph{International Journal of Theoretical Physics}, vol.~42, pp. 189--247, 2003.

\bibitem{boedihardjo2022privacy}
M.~Boedihardjo, T.~Strohmer, and R.~Vershynin, ``Privacy of synthetic data: A statistical framework,'' \emph{IEEE Transactions on Information Theory}, vol.~69, no.~1, pp. 520--527, 2022.

\bibitem{o1998information}
J.~A. O'Sullivan, R.~E. Blahut, and D.~L. Snyder, ``Information-theoretic image formation,'' \emph{IEEE Transactions on Information Theory}, vol.~44, no.~6, pp. 2094--2123, 1998.

\bibitem{bradlev1997use}
A.~Bradlev, ``The use of the area under the roc curve in the evaluation of machine learning algorithms,'' \emph{Pattern Recognition}, vol.~30, no.~7, pp. 1145--1159, 1997.

\bibitem{1381785}
L.~Araujo, L.~Sucupira, M.~Lizarraga, L.~Ling, and J.~Yabu-Uti, ``User authentication through typing biometrics features,'' \emph{IEEE Transactions on Signal Processing}, vol.~53, no.~2, pp. 851--855, 2005.

\bibitem{balabit}
K.~W. {Fülöp, Á.}, {Kovács, L.}, ``Balabit mouse dynamics challenge data set,'' 2016, \url{https://github.com/balabit/Mouse-Dynamics-Challenge/}, Last accessed on 03-01-2025.

\bibitem{antal2019user}
M.~Antal and L.~Denes-Fazakas, ``User verification based on mouse dynamics: a comparison of public data sets,'' in \emph{2019 IEEE 13th International Symposium on Applied Computational Intelligence and Informatics (SACI)}.\hskip 1em plus 0.5em minus 0.4em\relax IEEE, 2019, pp. 143--148.

\bibitem{shenchao}
C.~Shen, Z.~Cai, X.~Guan, Y.~Du, and R.~A. Maxion, ``User authentication through mouse dynamics,'' \emph{IEEE Transactions on Information Forensics and Security}, vol.~8, no.~1, pp. 16--30, 2013.

\bibitem{silverman2018density}
B.~W. Silverman, \emph{Density estimation for statistics and data analysis}.\hskip 1em plus 0.5em minus 0.4em\relax Routledge, 2018.

\bibitem{Heyman2001dataenough}
\BIBentryALTinterwordspacing
R.~E. Heyman, B.~R. Chaudhry, D.~Treboux, J.~Crowell, C.~Lord, D.~Vivian, and E.~B. Waters, ``How much observational data is enough? an empirical test using marital interaction coding,'' \emph{Behavior Therapy}, vol.~32, no.~1, pp. 107--122, 2001. [Online]. Available: \url{https://doi.org/10.1016/S0005-7894(01)80047-2}
\BIBentrySTDinterwordspacing

\bibitem{wortley2005dataenough}
\BIBentryALTinterwordspacing
A.~H. Wortley, P.~J. Rudall, D.~J. Harris, and R.~W. Scotland, ``How much data are needed to resolve a difficult phylogeny?: case study in lamiales,'' \emph{Systematic biology}, vol.~54, no.~5, pp. 697--709, October 2005. [Online]. Available: \url{https://academic.oup.com/sysbio/article-pdf/54/5/697/26543993/10635150500221028.pdf}
\BIBentrySTDinterwordspacing

\bibitem{kristen2013dataenough}
\BIBentryALTinterwordspacing
K.~D. Splinter, I.~L. Turner, and M.~A. Davidson, ``How much data is enough? the importance of morphological sampling interval and duration for calibration of empirical shoreline models,'' \emph{Coastal Engineering}, vol.~77, pp. 14--27, 2013. [Online]. Available: \url{https://www.sciencedirect.com/science/article/pii/S0378383913000495}
\BIBentrySTDinterwordspacing

\bibitem{staffuer1999guassian}
C.~Stauffer and W.~Grimson, ``Adaptive background mixture models for real-time tracking,'' in \emph{Proceedings. 1999 IEEE Computer Society Conference on Computer Vision and Pattern Recognition (Cat. No PR00149)}, vol.~2, 1999, pp. 246--252 Vol. 2.

\bibitem{silverman1986density}
B.~W. Silverman, \emph{Density Estimation for Statistics and Data Analysis}.\hskip 1em plus 0.5em minus 0.4em\relax Chapman and Hall/CRC, 1986.

\bibitem{kullback1951KL}
S.~Kullback and R.~A. Leibler, ``On information and sufficiency,'' \emph{The Annals of Mathematical Statistics}, vol.~22, no.~1, pp. 79--86, 1951.

\bibitem{shenchao2012continuous}
C.~Shen, Z.~Cai, and X.~Guan, ``Continuous authentication for mouse dynamics: A pattern-growth approach,'' in \emph{IEEE/IFIP International Conference on Dependable Systems and Networks (DSN 2012)}, 2012, pp. 1--12.

\bibitem{xujian2016mouseauth}
J.~Xu, M.~Li, F.~Zhou, and R.~Xue, ``Identity authentication method based on user's mouse behavior,'' \emph{Computer Science}, vol.~43, no.~2, pp. 148--154, 2016.

\bibitem{Bleha1990ComputerAccess}
S.~Bleha, C.~Slivinsky, and B.~Hussien, ``Computer-access security systems using keystroke dynamics,'' \emph{IEEE Transactions on Pattern Analysis and Machine Intelligence}, vol.~12, no.~12, pp. 1217--1222, 1990.

\bibitem{Obaidat1997Verification}
M.~S. Obaidat and B.~Sadoun, ``Verification of computer users using keystroke dynamics,'' \emph{IEEE Transactions on Systems, Man, and Cybernetics, Part B (Cybernetics)}, vol.~27, no.~2, pp. 261--269, 1997.

\bibitem{Krizhevsky2017ImageNet}
\BIBentryALTinterwordspacing
A.~Krizhevsky, I.~Sutskever, and G.~E. Hinton, ``Imagenet classification with deep convolutional neural networks,'' \emph{Communications of the ACM}, vol.~60, no.~6, pp. 84--90, June 2017. [Online]. Available: \url{https://doi.org/10.1145/3065386}
\BIBentrySTDinterwordspacing

\bibitem{Roy2023WCDCGAN}
A.~Roy, K.~Wong, and R.~C.-W. Phan, ``Attacking mouse dynamics authentication using novel wasserstein conditional dcgan,'' \emph{IEEE Transactions on Information Forensics and Security}, vol.~18, pp. 3622--3631, 2023.

\bibitem{Tan2019AdversarialAttacks}
Y.~X. Marcus~Tan, A.~Iacovazzi, I.~Homoliak, Y.~Elovici, and A.~Binder, ``Adversarial attacks on remote user authentication using behavioural mouse dynamics,'' in \emph{2019 International Joint Conference on Neural Networks (IJCNN)}, 2019, pp. 1--10.

\bibitem{gan}
Y.~X.~M. Tan, A.~Iacovazzi, I.~Homoliak, Y.~Elovici, and A.~Binder, ``Adversarial attacks on remote user authentication using behavioural mouse dynamics,'' 2019.

\bibitem{arjovsky2017wasserstein}
M.~Arjovsky, S.~Chintala, and L.~Bottou, ``Wasserstein generative adversarial networks,'' in \emph{Proceedings of the 34th International Conference on Machine Learning (ICML 2017)}, ser. ICML'17.\hskip 1em plus 0.5em minus 0.4em\relax JMLR.org, 2017, pp. 214--223.

\bibitem{mirza2014conditional}
M.~Mirza and S.~Osindero, ``Conditional generative adversarial nets,'' 2014.

\bibitem{jain2000filterbank}
A.~K. Jain, S.~Prabhakar, L.~Hong, and S.~Pankanti, ``Filterbank-based fingerprint matching,'' \emph{IEEE Transactions on Image Processing}, vol.~9, no.~5, pp. 846--859, 2000.

\bibitem{kinnunen2010overview}
\BIBentryALTinterwordspacing
T.~Kinnunen and H.~Li, ``An overview of text-independent speaker recognition: From features to supervectors,'' \emph{Speech Communication}, vol.~52, no.~1, pp. 12--40, 2010. [Online]. Available: \url{https://www.sciencedirect.com/science/article/pii/S0167639309001289}
\BIBentrySTDinterwordspacing

\bibitem{gupta2015keystroke_face}
A.~Gupta, A.~Khanna, A.~Jagetia, D.~Sharma, S.~Alekh, and V.~Choudhary, ``Combining keystroke dynamics and face recognition for user verification,'' in \emph{Proceedings of the 2015 IEEE 18th International Conference on Computational Science and Engineering (CSE)}, 2015, pp. 294--299.

\end{thebibliography}
\end{document}